\documentclass[12pt]{article}
\usepackage{graphicx}
\usepackage{epsfig,cite}
\newcommand{\be}{\begin{equation}}
\newcommand{\ee}{\end{equation}}
\newcommand{\bea}{\begin{eqnarray}}
\newcommand{\eea}{\end{eqnarray}}
\newcommand{\mP}{\mathcal{P}}
\newcommand{\mD}{\mathcal{D}}
\newcommand{\mF}{\mathcal{F}}
\newcommand{\mo}{\mathcal{O}}
\newcommand{\la}{\left\langle}
\newcommand{\ra}{\right\rangle}
\newcommand{\lc}{\left[}
\newcommand{\rc}{\right]}
\newcommand{\lp}{\left(}
\newcommand{\rp}{\right)}
\newcommand{\qb}{\bar{q}}
\newcommand{\bc}{\begin{center}}
\newcommand{\ec}{\end{center}}

\def\epm#1#2{\hbox{${\lower1pt\hbox{$\scriptstyle +~#1$}}
\atop {\raise1pt\hbox{$\scriptstyle -~#2$}}$}}

\begin{document}

\begin{flushright}
 
{\tt hep-ph/0401047} 
\\  UB--ECM--PF 04/01

\end{flushright}
 
\begin{center}

\vspace*{.6cm}

 {\bf{\Large  Neural network parametrization of spectral}} \\
{\bf{\Large functions from  hadronic tau decays }}\\ 
{\bf {\Large and determination of  QCD vacuum condensates}}

\vspace*{1.3cm}

{\bf Juan Rojo and
Jos\'e I. Latorre}

\vspace{0.5cm}

Departament d'Estructura i Constituents de la Mat\`eria, \\
Universitat de Barcelona, Diagonal 647, E-08028 Barcelona, Spain

\vspace*{2.5cm}
                                                                 
{\bf Abstract}

\end{center}
\noindent

The spectral function $\rho_{V-A}(s)$ is determined from ALEPH and OPAL 
data on hadronic tau decays using a neural network
parametrization trained to retain the full experimental information on errors,
their correlations and chiral sum rules: the DMO sum rule, the
first and second Weinberg sum rules and the electromagnetic mass
splitting of the pion sum rule. Nonperturbative 
QCD vacuum condensates can then be determined from
finite energy sum rules.
Our method minimizes all sources of theoretical 
uncertainty and bias  producing  an
estimate of the condensates which is 
independent of the specific finite energy sum rule
used. The results for the central values of the 
condensates $\la \mo_6\ra$ and  $\la \mo_8\ra$ are both negative.

\vfill 
\begin{flushleft} 
January 2004 
\end{flushleft}

\eject

\section{Introduction}
As the predictions of QCD become increasingly precise and more high quality
data is available, the theoretical uncertainties associated with the
analysis of the data are  often found to be dominant and thus
have come under increasing scrutiny. In this work
we continue previous efforts \cite{alphas,piccione}
to optimize the analysis of the 
information contained in the experimental data, 
taking into account errors and correlations, while introducing
the smallest possible theoretical bias. We consider the determination
of the QCD vacuum condensates $\la \mo_6 \ra,\la \mo_8 \ra$ and
higher
dimensional condensates,  
which in principle
can be extracted in a theoretically solid and experimentally clean way
from the hadronic decays of the tau lepton. In practice, however,
the situation is far from satisfactory, as revealed by the lack of stability
of the value of the nonperturbative condensates obtained from this process by
different procedures.

The main source of difficulties can be traced to the fact that conventional
extractions of the QCD vacuum condensates
 from hadronic tau decays involve convolutions
of the difference  $v_1(s)-a_1(s)$ of the isovector vector and
axial vector spectral functions, which is a purely nonperturbative quantity,
that does not converge to the perturbative result (these spectral
functions are degenerate within perturbation theory) at energies 
$s\le M_{\tau}^2$ and moreover has large uncertainties in the high $s$
region. To obtain a reliable
value for the nonperturbative condensate, some convergence method must be
applied, implying that the error on the condensates gets tangled with 
uncertainties within the theoretical assumptions and subject therefore
to a variety of sources of theoretical bias. This is a consequence of
the fact that the kinematics of the hadronic tau decays constrains
the range of energies in which we can evaluate the spectral functions. 
The main difficulty is that any  method to estimate non-perturbative
condensates exploits the shape
of the spectral functions near and beyond the boundary of the
region where the data is available. Final results are then
subject to systematics errors associated to the extrapolation
of data as well as the way global theoretical constraints, {\it e. g.}
Weinberg sum rules, are imposed.

In this paper, we approach the determination of 
nonperturbative condensates, in a way
which tries to bypass these difficulties, by combining two techniques. 
First, a novel {\it bona fide} method to take
into account experimental errors
was proposed and implemented in the 
context of analysis of Deep Inelastic Scattering data
to produce a probability measure in the space of deep inelastic structure
functions by means of neural networks \cite{structure}. 
Here, we adapt this method to the parametrization of
spectral functions. The second
technique we use refines the training
of neural networks so as to 
implement the constraints that
represent the QCD chiral sum rules in our
neural network parametrization of the spectral function $v_1(s)-a_1(s)$.

The representation of the probability density given in Ref. \cite{structure}
takes the form of a set of neural networks, trained on an ensemble of Monte 
Carlo replicas of the experimental data, which reproduce their probability
distribution. The parametrization is unbiased 
in the sense that neural networks do not 
rely on the choice of an specific functional form, and it interpolates
between data points, imposing smoothness constraints in a controllable 
way. Information on experimental errors and correlations is
incorporated in the Monte Carlo sample. Errors on
physical quantities and correlations between them can then be determined
without the need of linearized error propagation. Our
final parametrization 
combines all the available experimental information, as well as constraints
from different convolutions of the data, i.e. chiral sum rules, must
verify. In this way statistical errors can be estimated and 
the loss of accuracy due to the  extrapolation outside
the kinematical region where the data is available is also analyzed.

Hence, we can obtain a determination of the QCD vacuum condensates 
which is unbiased with respect to the parametrization of the
spectral function and the error and correlations propagation.
We also try to keep under control all sources of
uncertainty related to the method of analysis,
and estimate their contribution to the total error.
This gives us a  determination of the nonperturbative condensates, and
simultaneously illustrates the power of a method of analysis based
on direct knowledge of a probability density in a space of functions.

This paper is organized as follows: in section 2 we review
theoretical tools used in the analysis of non-perturbative
effects in hadronic tau decays; in section 3 we present the 
experimental data that is used in our analysis and in section 4 we introduce
the neural network parametrization of spectral functions. Section 5
contains the details and results of our extractions of the vacuum
condensates: we explain our choice of training parameters, our
error estimations and the consistency tests that we performed; finally
section 6 summarizes our conclusions.

\section{Spectral functions and hadronic tau decays}

The tau particle is the only  lepton massive enough to
decay into hadrons. Already before its discovery, it was predicted to be
important for the study of hadronic physics \cite{tsai}, a study
that has been performed extensively at the LEP accelerator.
Its semileptonic decays are therefore an ideal tool for 
studying the hadronic weak currents under clean conditions, both
theoretically and experimentally, thanks to the high quality data
from LEP. 
In this first section we will briefly introduce
the theoretical foundations that form the basis of the QCD analysis
of hadronic tau decays. There exists a huge
literature on tau hadronic physics to which the interested reader
is directed (see for instance Ref. \cite{tauthesis} and references therein).

Spectral functions are the observables that give access to the inner 
structure of hadronic tau decays. As parity is maximally violated in
$\tau$ decays, the spectral functions will have both vector
and axial vector contributions. As spontaneous chiral symmetry
breaking is a nonperturbative phenomena, this spectral functions
are degenerate in perturbative QCD with massless light quarks, so 
any difference
between vector and axial-vector spectral functions is necessarily 
generated by non-perturbative dynamics, that is, long distances 
resonance phenomena, being the most relevant the $\rho(770)$ and
the $a_1(1260)$ in the vector and axial vector channels respectively. 
Therefore, the difference of these spectral functions is generated entirely 
from nonperturbative QCD dynamics, and provides a laboratory
for the study of these perturbative contributions, which have resulted
to be small and therefore
difficult to measure in other processes where the 
perturbative contribution dominates. An accurate extraction
of the QCD vacuum condensates is not only important by itself but
also has many important phenomenological applications, for example
in the evaluation of matrix elements in weak decays \cite{buras}.
 
The ALEPH collaboration at LEP measured 
 \cite{exp1,exp2} these
spectral functions from hadronic tau decays with great accuracy, providing
an excellent source of precision analysis of nonperturbative
effects. As it is well known \cite{pich}, the basis of the comparison of
theory with data is the fact that unitarity and analyticity connect
the spectral functions of hadronic tau decays to the imaginary part of 
the hadronic vacuum polarization,
\be
\Pi_{ij,U}^{\mu\nu}(q)\equiv \int d^4x~e^{iqx}\la 0|T\lp U^{\mu}_{ij}(x)
 U^{\nu}_{ij}(0)^{\dag}
\rp|0\ra \ ,
\ee
of vector $ U^{\mu}_{ij}\equiv V^{\mu}_{ij}=\qb_j\gamma^{\mu}q_i$
or axial vector 
$ U^{\mu}_{ij}\equiv A^{\mu}_{ij}=\qb_j\gamma^{\mu}\gamma_5q_i$
color singlet quark currents in corresponding quantum states. 
After Lorentz decomposition is used to separate the correlation function into 
its $J=1$ and $J=0$ components,
\be
\Pi_{ij,U}^{\mu\nu}(q)=\lp-g^{\mu\nu}q^2+q^{\mu}q^{\nu}\rp\Pi^{(1)}_{ij,U}(q^2)
+q^{\mu}q^{\nu}\Pi^{(0)}_{ij,U}(q^2) \ ,
\ee
for non-strange quark currents one 
identifies
\be
\mathrm{Im} \Pi^{(1)}_{\bar{u}d,V/A}(s)=\frac{1}{2\pi}v_1/a_1(s) \ .
\ee
This relation allows us to implement all the technology of QCD vacuum 
correlation functions to hadronic tau decays, and provides the 
basis of the comparison of theory with data.

The basic tool to study in a systematic way the power corrections
introduced by nonperturbative dynamics is the operator product 
expansion. 
Since the approach of Ref. \cite{svz}, the operator product expansion (OPE)
has been used to perform calculations with QCD on the ambivalent
energy regions where nonperturbative effects come into play
but still perturbative QCD is relevant. In general, the OPE of a two
point correlation function $\Pi^{(J)}(s)$ takes the form \cite{pich}
\be
\Pi^{(J)}(s)=\sum_{D=0,2,4,\ldots}\frac{1}{(-s)^{D/2}}\sum_{\mathrm{dim}
\mo=D}C^{(J)}(s,\mu)\la\mo(\mu)\ra \ ,
\ee
where the arbitrary parameter $\mu$ separates the long distance 
nonperturbative effects absorbed into the vacuum expectation elements
$\la\mo(\mu)\ra$, from the short distance effects which are included
in the Wilson coefficient $C^{(J)}(s,\mu)$. The operator of dimension
$D=0$ is the unit operator (perturbative series), and  we are interested in
the dimension $D\ge 6$ operators. 
What is relevant for us is that  $D=6$ 
is the first non-vanishing non-perturbative contribution, in the limit of
massless light quarks, to the $v_1(s)-a_1(s)$ spectral function and,
moreover, it has been shown to be the dominant one. 
This dominant contribution
carries non-trivial four-quark dynamical effects of the form $\qb_i
\Gamma_1q_j\qb_k\Gamma_2q_l$. Additional contributions from a mixed quark
gluon condensate as well as a triple gluon condensate are assumed to
be small.
Therefore, 
this spectral function should provide a source for a clean
extraction of the value of the nonperturbative contributions.

Finally, we review the important paper that QCD chiral sum rules
play in the analysis of this process.
Sum rules have always been an important tool for
studies of non-perturbative aspects of QCD, and have been applied to a wide
variety of processes, from Deep Inelastic Scattering to Heavy Quark systems
\cite{derafael},\cite{srnarison}. Now we will review one of the 
classical examples of low energy QCD sum rules, 
the chiral sum rules. 
The application of chiral symmetry together with the optical theorem 
leads to low energy sum rules
involving the difference of vector and axial vector spectral functions,
\be
\rho_{V-A}(s)\equiv v_1(s)-a_1(s) \ .
\ee
These sum rules are dispersion relations 
between real and absorptive parts of a two point correlation function
that transforms symmetrically under $SU(2)_{L}\otimes SU(2)_{R}$ in the case
of non strange currents. Corresponding integrals are
the Das-Mathur-Okubo sum rule \cite{dmo}
\be
\label{dmo}
\frac{1}{4\pi}\int_0^{s_0\to\infty}ds\frac{1}{s}
\rho_{V-A}(s)=\frac{f_{\pi}^2<r_{\pi}^2>}{3}
-F_A \ ,
\ee
as well as the first and second Weinberg sum rules (WSR) \cite{weinberg}
\be
\frac{1}{4\pi^2}\int_0^{s_0\to\infty}ds\rho_{V-A}(s)=f_{\pi}^2 \ ,
\label{wsr1}
\ee
\be
\int_0^{s_0\to\infty}dss\rho_{V-A}(s)=0 \ ,
\label{wsr2}
\ee
where in eq. (\ref{wsr1}) the RHS term comes from the integration of the 
spin zero axial contribution, which for massless non-strange quark 
currents consists exclusively of the pion pole. Finally, there is the
chiral sum rule associated with
 the electromagnetic splitting of the pion masses \cite{empion},
\be
\label{empionmass}
\frac{1}{4\pi^2}\int_0^{s_0\to\infty}dss\ln \frac{s}{\lambda^2}\rho_{V-A}(s)=
-\frac{4\pi f_{\pi}^2}{3\alpha}(m_{\pi^{\pm}}^2-m_{\pi^0}^2) \ ,
\ee
where $f_{\pi}=(92.4\pm0.3)$ MeV \cite{pdg} obtained from the decays
$\pi^-\to \mu^-\bar{\nu}_{\mu}$ and $\pi^-\to \mu^-\bar{\nu}_{\mu}\gamma$, 
$F_A=0.0058\pm0.00008$ is the pion axial vector 
form factor\footnote{Note that our definition 
of $F_A$ agrees with that of Ref. \cite{exp2} 
but differs by a factor of $1/2$ from that given in Ref. \cite{pdg}} 
obtained from the radiative
decays $\pi^-\to l^-\bar{\nu}_{l}\gamma$ and $\la r_{\pi}^2\ra=
(0.439\pm0.008)$ fm$^2$ is the pion charge radius squared. From now on
these four chiral sum rules will be denoted by SR1, SR2, SR3 and
SR4 respectively. It could be argued that as long that these chiral
sum rules are taken in the chiral limit, the value of the pion
decay constant should be the chiral limit value, $f\sim 0.94f_{\pi}$
\cite{qpt}. However, as long as the experimental data consists of
real pions, we consider that it is more reasonable to use the real
world value for the pion decay constant.

When switching quark masses on, only the first WSR remains valid while
the second breaks down due to contributions from the difference of 
non-conserved vector and axial vector currents of order $m_q^2/s$ 
leading to a quadratic divergence of the integral. This is not 
numerically relevant
in our analysis because we deal with finite energy sum rules, and in this
case the contribution from non-zero quark masses is negligible.

The QCD vacuum condensates can be determined by virtue of the dispersion
relation from another sum rule, that is, a
 convolution of the $\rho_{V-A}(s)$ spectral function 
with an appropriate weight function. Let us define the operator product
expansion of the chiral correlator in the following way
\be
\label{chiralope}
\Pi(Q^2)|_{V-A}=\sum_{n=1}^{\infty}\frac{1}{Q^{2n+4}}C_{2n+4}(Q^2,\mu^2)
\left\langle \mathcal{O}_{2n+4}(\mu^2)\right\rangle
\equiv \sum_{n=1}^{\infty}\frac{1}{Q^{2n+4}}
\left\langle \mathcal{O}_{2n+4}\right\rangle \ .
\ee
The Wilson coefficients, including radiative corrections, are
absorbed into the nonperturbative vacuum expectation values, to
facilitate comparison with the current literature.  
The analytic structure of the $\Pi$ is subject to
the dispersion relation
\be
\Pi_{V-A}(Q^2)=\int_0^{\infty}\mathrm{d}s\frac{1}{s+Q^2}\frac{1}{\pi}
\mathrm{Im}\Pi_{V-A}(s)\ .
\ee
Condensates of arbitrary dimension are simply given by
\be
\label{condenconv}
\left\langle\mathcal{O}_{2n+2}\right\rangle= 
(-1)^n\int_0^{s_0}dss^n\frac{1}{2\pi^2}\lp v_1(s)-a_1(s) \rp \ ,\quad
n\ge 2
\ ,
\ee
which,if the asymptotic regime has reached, should be independent
of the upper integration limit for large enough $s_0$. 
As can be seen from the experimental data, errors
in the large $s$ region are very important, so large errors are expected
in the evaluation of the condensates.
The analysis of these sources of  errors is one of our main goals
in the present
 analysis, which will be obtained thanks to the natural capability
of neural networks of smooth interpolating while implementing all the
experimental information on errors and correlations. 

\subsection{Finite energy sum rules}
As long as all previous integrals have to be cut at some finite 
energy $s_0\le M_{\tau}^2$, since no experimental information
on $v_1(s)-a_1(s)$ is available above $M_{\tau}^2$, we must perform
a truncation that competes with all other sources of statistical and 
systematic errors, introducing a theoretical bias which is difficult
to estimate. Many techniques have been developed to deal
with this finite energy integrals, leading to the so-called 
Finite Energy Sum Rules (FESR). The paradigmatic example is the calculation
of spectral moments \cite{diberder}, 
that is weighted integrals over spectral functions. Choosing 
appropriate weights allows to extract the maximum information possible
from the experimental data while minimizing the contribution from
the region with larger errors. This techniques allow
a comparison of the same quantity evaluated on one side with
experimental data and on the other side with theoretical input,
basically the Operator Product Expansion with perturbative QCD corrections.
The general expression that takes advantage of the analyticity properties
of the chiral correlators is given by
\be
\label{analitic}
\int_{0}^{s_0}ds~W(s)~\mathrm{Im}\Pi^{(J)}_{V-A}(s)=\frac{1}{2i}
\oint_{|s|=s_0}ds~W(s)~ \Pi^{(J)}_{V-A}(s) \ ,
\ee
where $W(s)$ is an analytic function and $s_0$ is large enough for the
OPE series to converge. The LHS of eq. (\ref{analitic}) can be evaluated
using the experimental input from spectral functions as determined
in hadronic tau decays, while the RHS can be evaluated using the OPE
representation of the chiral correlator. Finally, a fit is
performed to extract the OPE parameters from the experimental
data on spectral functions. 

A common hypothesis in the majority of this kind of analysis is
that the difference of the OPE representation for the chiral correlator
from the full expression,
\be
R[s_0,W]\equiv\frac{-1}{2\pi i}\int_{|s|=s_0}ds\lp \Pi_{V-A,OPE}(s)
-\Pi_{V-A}(s) \rp
W(s) \ ,
\ee can be neglected. This quantity is a measure of the 
OPE breakdown, also known as duality violation\footnote{
For a review of the current theoretical status of the quark-hadron
duality violations, see Ref. \cite{shifman}}. It is necessary 
to take into account that $\Pi_{V-A,OPE}$ fails at least in 
some region of the integration contour. 
This was shown in Ref. \cite{poggio}, where it was demonstrated that
the OPE representation breaks down near the timelike real $s$ axis for
insufficiently large $s_0$.
The neglect of the duality violation component of the OPE is a key dynamical
assumption and there exists several strategies to minimize its impact, 
as working with {\it duality points} \cite{peris} 
or using pinched Finite Energy
Sum Rules \cite{cirigliano}, with polynomial weights 
that vanish at the upper integration limit. 
All these techniques yield different although compatible 
values for the $\la \mo_6\ra$, whereas non-compatible
results are obtained for higher condensates.

Other types of finite energy
sum rules  have been used to extract the values of the
condensates and other phenomenologically relevant related quantities. 
Borel sum rules and Gaussian sum rules \cite{ioffe} take advantage of certain 
combination of the condensates that theoretically optimize the
accuracy of the extraction. Inverse moment sum rules \cite{davier} 
techniques
make a connection between the phenomenological parameters of the 
QCD effective Lagrangian and the nonperturbative condensates. In section 
5 we will compare our extraction of the condensates to those
obtained with all these methods and argue why ours has a reasonable
control of the different theoretical uncertainties.

Our approach is different with respect to previous determinations
of nonperturbative condensates. First of all, we use the smooth
interpolation
of the neural network to extend the range of integration, so that
our determination of the condensates corresponds to the
asymptotic energy region $s_0\to\infty$. The second point is that
the training method allows the incorporation of the chiral sum rules
to the neural network parametrization of the spectral function, and
therefore
the specific weight used to determine the condensate turns out
not to be relevant: different choices of weights differ by chiral
sum rules that are already verified our parametrization. 
Therefore, the final results for the non-perturbative
condensates that emerge from
the neural network parametrization of the spectral function
$v_1(s)-a_1(s)$ are determined by
\be
\label{condenconv2}
\left\langle\mathcal{O}_{2n+2}\right\rangle= 
(-1)^n\int_0^{\infty}dss^n\frac{1}{2\pi^2}\lp v_1(s)-a_1(s) \rp \quad n\ge 2
\ .
\ee
In the next sections will be argued why this choice is the
most reasonable one, showing that all  relevant 
constraints are verified.

\section{Experimental data}
Since the relevant spectral function for the determination of the
condensates is the $v_1(s)-a_1(s)$ spectral function, we need a
simultaneous measurement of the vector and axial-vector spectral
functions. Data from the ALEPH Collaboration \cite{exp2}, 
\cite{exp1} and from the OPAL collaboration \cite{opaldata} will be used,
 which provide a simultaneous determination of the vector
and axial vector spectral functions in the same kinematic region and
also provide the full set of correlated uncertainties for these measurements.
Although the ALEPH data is of a higher quality due to the smaller
errors, see fig.  (1), 
the input from OPAL is complementary and will provide a
cross check for our extractions of the nonperturbative condensates. 
There exists additional data on spectral functions coming from
electron-positron annihilation, but their quality is lower than the
data from hadronic tau decays and will be ignored here.

ALEPH 
experimental data consists on the invariant square-mass
spectra for both the vector+axial vector and vector-axial vector
components, that are related to the spectral functions
by a kinematic factor and a branching ratio
normalization
\be
v_1(s)/a_1(s)\equiv\frac{M^2_{\tau}}{6|V_{ud}|^2S_{EW}}\frac{B(\tau^-\to
V^-/A^-\nu_{\tau} )}{B(\tau^-\to e^-\bar{\nu}_e\nu_{\tau}
)}\frac{dN_{V/A}}{N_{V/A}ds}\left[\left(1-\frac{s}{M_{\tau}^2}
\right)\left(1-\frac{2s}{M_{\tau}^2}\right)\right]^{-1} \ .
\ee  

\begin{figure}
\begin{center}
\includegraphics[scale=0.28,angle=-90]{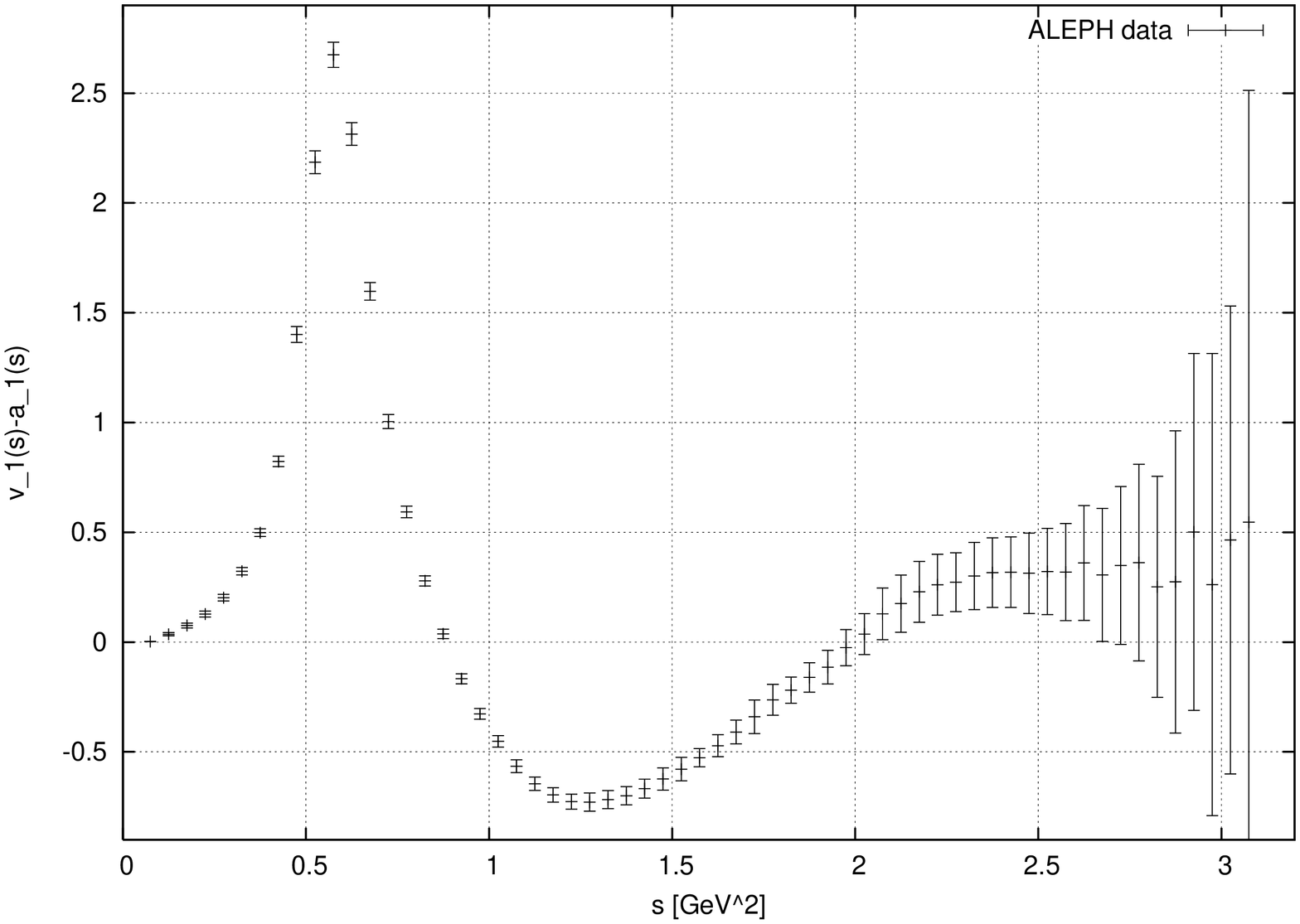}
\includegraphics[scale=0.28,angle=-90]{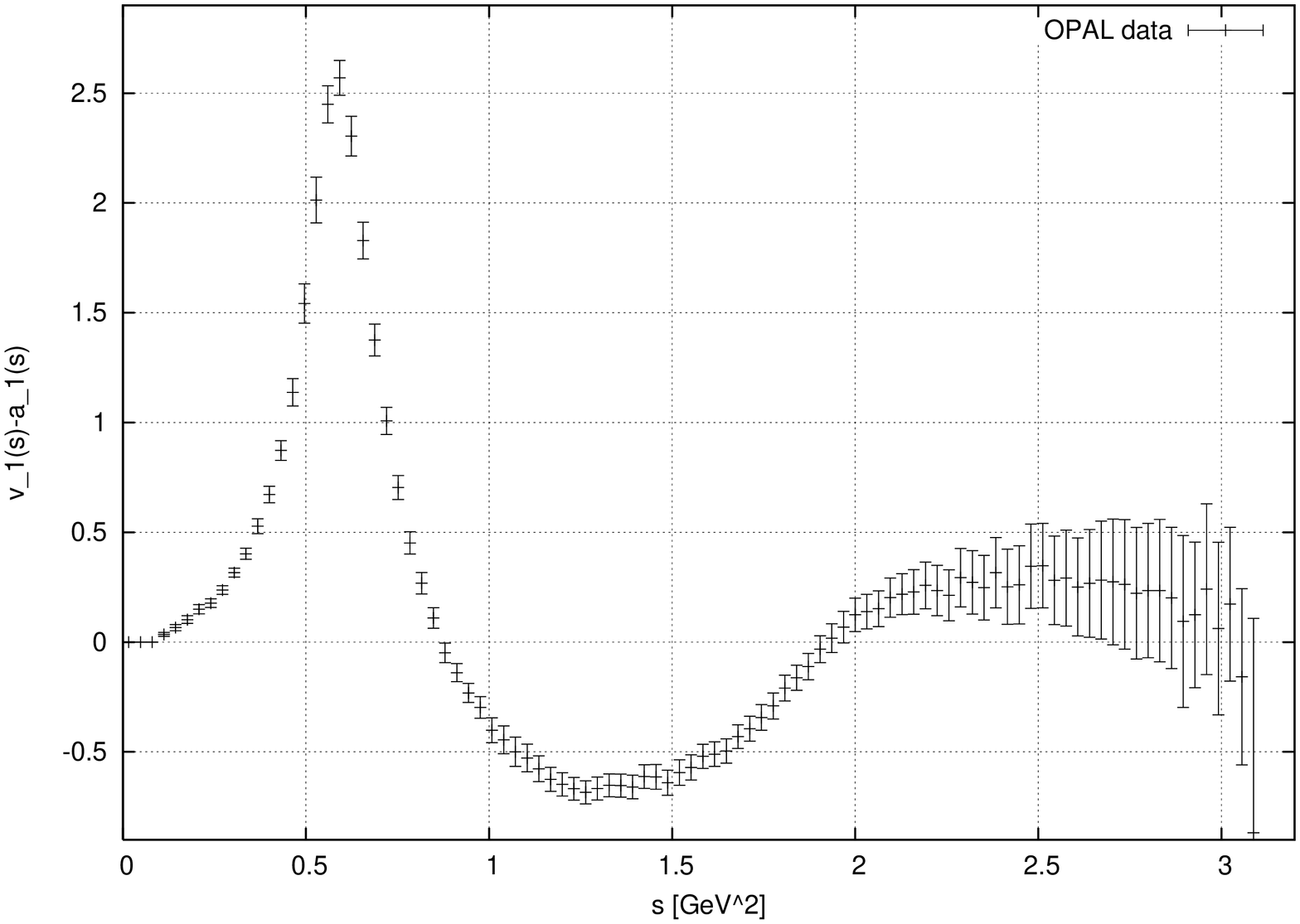}
\label{specdata}
\caption{}{Experimental data for $v_1(s)-a_1(s)$ spectral function from
  the ALEPH (left) and OPAL (right) collaborations. Note that the
  errors are 
smaller in the ALEPH data but OPAL central values are nearer to
the expected zero value at large $s$.}
\end{center}
\end{figure}

Altogether our parametrization is based on $N_{dat}=$ 61 experimental 
points for ALEPH, although the full experimental data consists in
70 points uniformly distributed between 0 and 3.5 GeV$^2$, because only 
points with $s\le M_{\tau}^2=3.16$ GeV$^2$ are physically
meaningful, and before this kinematic threshold is reached the
invariant mass-squared spectrum vanishes due to phase-space 
suppression. For OPAL the data sample is a bit larger, $N_{dat}=97$, with
the same restrictions as in the ALEPH case. 
 Henceforth, 
$\rho^{(exp)}_{V-A,i}$ will denote the $i$-th data point $\rho_{V-A}(s_i)\equiv
v_1(s_i)-a_1(s_i)$. Figure (\ref{specdata}) 
shows the experimental data
used together with diagonal errors.

Note that errors are  small in the low and middle $s$
regions and that they become larger as we approach the tau mass threshold.
The last points are almost zero in the invariant mass spectrum, and
are only enhanced in the spectral functions due to the large
kinematic factor for $s$ near $M_{\tau}^2$, so  special
care must be taken with the physical relevance of these points. 

It is clear that the vanishing of the spectral function
is not reached for $s\le M_{\tau}^2$ and must be enforced
artificially on the parametrization we are constructing,
that is  we must device a technique to impose the 
asymptotic constraint that at high $s$ this spectral function vanishes. 
The method we use
takes advantage of the smooth, unbiased interpolation
capability of the neural network: artificial points are added to
the data set with adjusted errors in a region where 
$s$ is high enough that the $\rho_{V-A}(s)$ spectral function
should vanish. Once these artificial points are
included, in a way to be discussed later,
 the neural network will smoothly interpolate between the
real and artificial data points, also taking into account
 the constraints of the sum  rules, as explained below.

The experimental data points are highly correlated, because the majority
of the covariance matrix is composed of nonzero entries, so it is therefore
crucial to take into account all their correlations, which are
specially relevant in the high $s$ region. 
This is important because this region dominates the
sum rule, eq. (\ref{condenconv}), that determines the vacuum condensates.
As we shall discussed shortly, 
correlated errors are incorporated  as a measure on the space 
of neural network 
parametrizations of the spectral functions using 
 Monte Carlo
statistical replicas of the experimental data.

\section{Neural network parametrization}
Ideally, a parametrization of spectral functions must incorporate
all the information contained in the experimental measurements, i.e. their
central values, their statistical and systematic errors and their 
correlations, furthermore, it must interpolate between them without
introducing any bias. We will follow the method of Ref. \cite{structure}, 
where an unbiased extraction of the probability measure in the space
of structure functions of deep-inelastic scattering is performed, based
on a coordinated use of Monte Carlo generation of data and neural network fits.

\subsection{Probability measure in the space of spectral functions}
The experimental data gives us a probability measure in an $N_{dat}$
dimensional space, assumed to be multigaussian. In order to extract from it
a parametrization of the desired structure function we must turn this measure
into a measure $\mP\lc \rho_{V-A} \rc$ in a space of functions. 
Once such a measure is constructed, the expectation value of any observable
$\mF\lc \rho_{V-A}(s)\rc $ can be found by computing the weighted average
\be
\label{wa}
\la \mF\lc \rho_{V-A}(s)\rc \ra=\int\mD\rho_{V-A}~\mF\lc \rho_{V-A}(s)
\rc~\mP\lc\rho_{V-A}\rc.
\ee
Errors and correlations can also be obtained from this measure,
by considering higher moments of the same observable with respect to the
probability distribution.

The determination of an infinite-dimensional measure from a finite set
of data points is an ill-posed problem, unless one introduces further 
assumptions. In the approach of Ref. \cite{structure}, neural networks
are used as interpolating functions, so that the only assumption is the 
smoothness of the spectral function. Neural networks can fit any
continuous function through a suitable training; smoother functions require
a shorter training and less complex networks. Hence, an ideal degree
of smoothness can be established on the basis of a purely statistical
criterion without the need for further assumptions.

\subsection{Fitting strategy}
The construction of the probability measure is done in two steps: first, a set 
of Monte Carlo replicas of the original data is generated. This gives a 
representation of the probability density $\mP\lc\rho\rc$ at points $(s_i)$ 
where data is available. Then a neural network is fitted to each replica. 
The ensemble of neural networks gives a representation of the probability
density for all $s$: when interpolating between data the uncertainty
will be kept under control by the smoothness constraint, but it will
become increasingly more sizable when extrapolating away from the data region.

The $k=1,\ldots,N_{rep}$ replicas of the data are generated as
\be
\label{rep}
\rho_{V-A,i}^{(art)(k)}=\rho_{V-A,i}^{(exp)}+r_i^{(k)}\sigma_i \ ,
\ee
where $\rho_{V-A,i}^{(exp)}=\rho_{V-A}(s_i)$ are 
the original data, $\sigma_i$ is
the diagonal error, and $r_i^{(k)}$ are univariate gaussian random
numbers whose correlation matrix equals that of the experimental data.
The fact that the correlation matrix of the  $r_i^{(k)}$ equals
that of the experimental data is crucial to retain all the experimental
information in our treatment.
Then a set of $N_{rep}=1000$ replicas of this form is generated, and 
is verified that the central values, errors and correlations of the 
original experimental data are well reproduced by taking the relevant 
averages over a sample of this size. A explained above,
 the asymptotic constraint that
$\rho_{V-A}(s\to\infty)=0$ has been 
implemented by adding a number of artificial
data points with adjusted errors.

To verify that the central values, errors and correlations of the
original experimental data are well reproduced, we can define statistical
estimators that measure the deviations from the original correlations. 
A suitable one is the scatter correlation, which measure the deviations
of the averages over the replica set from the original experimental values,
and are defined as follows for the central value
\be
\label{scatterrep}
r[\rho_{V-A}^{(art)}]=\frac{\la\rho_{V-A}^{(exp)}\la
\rho_{V-A}^{(art)}\ra_{rep}
\ra_{dat}-
\la\rho_{V-A}^{(exp)}\ra_{dat}\la\la
\rho_{V-A}^{(art)}\ra_{rep}
\ra_{dat}
}{\sigma_{s}^{(art)}\sigma_{s}^{(exp)}} \ ,
\ee
where the scatter variances are defined by
\be
\sigma_s^{(exp)}=\sqrt{\la\left(\rho_{V-A}^{(exp)}\right)^2\ra_{dat}-
\left(\la\rho_{V-A}^{(exp)}\ra_{rep}\right)^2} \ ,
\ee
\be
\sigma_s^{(art)}=\sqrt{\la\left(\la\rho_{V-A}^{(exp)}\ra_{rep}
\right)^2 \ra_{dat}-
\left(\la\la\rho_{V-A}^{(art)}\ra_{rep}\ra_{dat}\right)^2} \ ,
\ee
and similarly for the diagonal errors and the correlations.
In table \ref{reptesttable} we show the scatter correlations
for the central values, the errors and the correlations. We observe
that we need $N_{rep}=100$ replicas to maintain the correlations
of the original data, which is the main purpose of our analysis. We
have
checked that increasing the number of training replicas does not
decrease
the errors in the extraction of the condensate further, meaning that
we have reached a faithful representation of errors.

\begin{table}[t]  
\begin{center}  
\begin{tabular}{cccc} 
\multicolumn{4}{c}{$\rho_{V-A}(s)$}\\   
\hline
$N_{rep}$ & 10 & 100 & 1000 \\
\hline  
$r [\rho_{V-A}(s)^{(art)}]$ 
  & 0.9803  & 0.9997 & 0.9998 \\
\hline  
$r [\sigma^{(art)} ]$ 
  & 0.9894 & 0.9992 & 0.99994 \\
\hline
$r [\rho^{(art)}]$ 
  & 0.61 & 0.955 & 0.9956 \\
\hline
\end{tabular}
\end{center}
\caption{Comparison between experimental and Monte Carlo 
generated artificial data for
the $\rho_{V-A}(s)$ spectral function. Note that the scatter
correlation $r$, defined in eq. (\ref{scatterrep}), for $N_{rep}=100$
is
already
very close to 1 for all statistical estimators.}
\label{reptesttable}
\end{table}

Each set of generated data is fitted by an individual neural network. A 
neural network \cite{netrev},\cite{phd} is a function of a number of 
parameters,
which fix the strength of the coupling between neurons and the threshold
of activation of each neuron. The architecture of the network has
been  chosen to be 1-4-4-1, small enough to avoid overlearning
and large enough to capture the non-linear structure of 
experimental data.
The networks that we use are
multilayer feed-forward neural networks 
constructed according to the following recursive relation
\be
 \xi^{(l)}_i=g\lp h_i^{(l)}\rp \ ,
\ee
\be
h_i^{(l)}=\sum_{j=1}^{n_{l-1}}\omega_{ij}^{(l)}\xi_{j}^{(l-1)}-\theta_i^{(l)}
\ ,
\ee
where $\omega_{ij}^{(l)}$ is the weight, the strength of the connection
between two neurons, $\theta_i^{(l)}$ are the thresholds of each neuron,
 $\xi^{(l)}_i$ is the activation state of each neuron and $g$ is the
activation function of the neurons.
 
We divide the training of the neural network in two epochs.
 In a first epoch, the training method
is done by
backpropagation, where the parameters of the network 
are fitted by 
minimizing the error function:
\be
\label{err}
E^{(k)}_{\mathrm{err}}=\sum_{i=1}^{N_{dat}}\frac{\lp\rho_{V-A,i}^{(art)(k)}
-\rho_{V-A,i}^{(net)(k)}\rp^2}{\sigma_i^{(exp)^2}} \ ,
\ee
where $\rho_i^{(net)(k)}$ is the prediction of the $i-$th data point from
the net trained on the $k-$th replica of the data. A more detailed
review of neural networks learning techniques is presented in the
appendix \ref{nntech}. 

In a second training epoch, a different training technique called
genetic 
algorithms training is used to 
implement the constraints from the sum rules. As explained below and in the
appendix \ref{nntech}, this technique allows us to implement in our
training non-local constraints, as convolutions of the neural
network output, with  adjusted weights so that the chiral sum
rules control the neural network interpolation in the data
region where errors are greater. 
The error function eq. (\ref{err}) is
modified by adding a contribution proportional to the difference
of the chiral sum rules, evaluated with the output of the trained
networks, and their theoretical values, that is
\be
\label{err2}
E_{\mathrm{tot}}=E_{\mathrm{err}}+E_{\mathrm{sr}}=E_{\mathrm{err}}+
\sum_{i=\mathrm{sr}_1}^{\mathrm{sr}_4}w_{\mathrm{sr}_i}\lp
\int_0^{s_0} \mathrm{d}s f_i\lp\rho_{V-A}(s)\rp-A_i\rp^2 \ ,
\ee
where $w_{\mathrm{sr}_i}$ is the relative weight of each sum rule
 and $A_i$ is the theoretical value of
the corresponding sum rule, eqns. 
(\ref{dmo}-\ref{empionmass}). 
We note that this definition introduces a new set of, in 
principle, arbitrary parameters, that is the relative weights of the sum rules.
As explained below, these are determined by stability criteria,  
demanding that the contribution from $E_{\mathrm{sr}}$ is similar
to that of $E_{\mathrm{err}}$ and that the final result
is not sensitive to the specific values of these parameters.

The basic idea of the genetic algorithms training, also known as
natural selection training, works as follows. 
The training is divided in generations. For
 each generation the parameters
of the network (weights and thresholds) are arranged to form a chain, 
called the ADN chain. This chain is replicated many times, creating
a population of identical individuals. Later, random mutations are
applied to each individual, where by mutation we mean a small change
in one of the bits of his ADN chain.
Then, the error function
associated with each mutated individual is computed, which implies 
passing back
the ADN bits to their original status of weights and thresholds and
calculating the output of this new network. Only the best individuals
are kept
while discarding the rest, mimicking natural selection.
 This method provides a suitable
technique to implement the effect of the chiral sum rules on our
neural network training. Note that this technique leads to an important
increase on the computing time, due to the fact that the chiral sum rules
must be numerically evaluated many times each generation.

The main advantage of genetic algorithms is that they allow neural networks
to learn from error functions that may be as complicate as 
making impossible the use of backpropagation training. Furthermore,
genetic algorithms can be proven to search efficiently the
parameter space of  solutions, exploring exponentially 
many more times reasonable outputs as compare to manifestly
wrong ones. Genetic algorithms can also handle the training
of very large neural networks.

The parametrization obtained by means of the genetic algorithm
training
is represented in
 figure (2) where the output of the trained neural network,
without and with the inclusion of
the
chiral sum rules,
is compared with the experimental data points together with the corresponding
statistical errors. It is clear that the effect of the chiral
sum 
rules on the trained neural network output is forcing it to reach
faster
the asymptotic behavior of the spectral function $\rho_{V-A}(s)$.

\begin{figure}
\begin{center}
\includegraphics[scale=0.33,angle=-90]{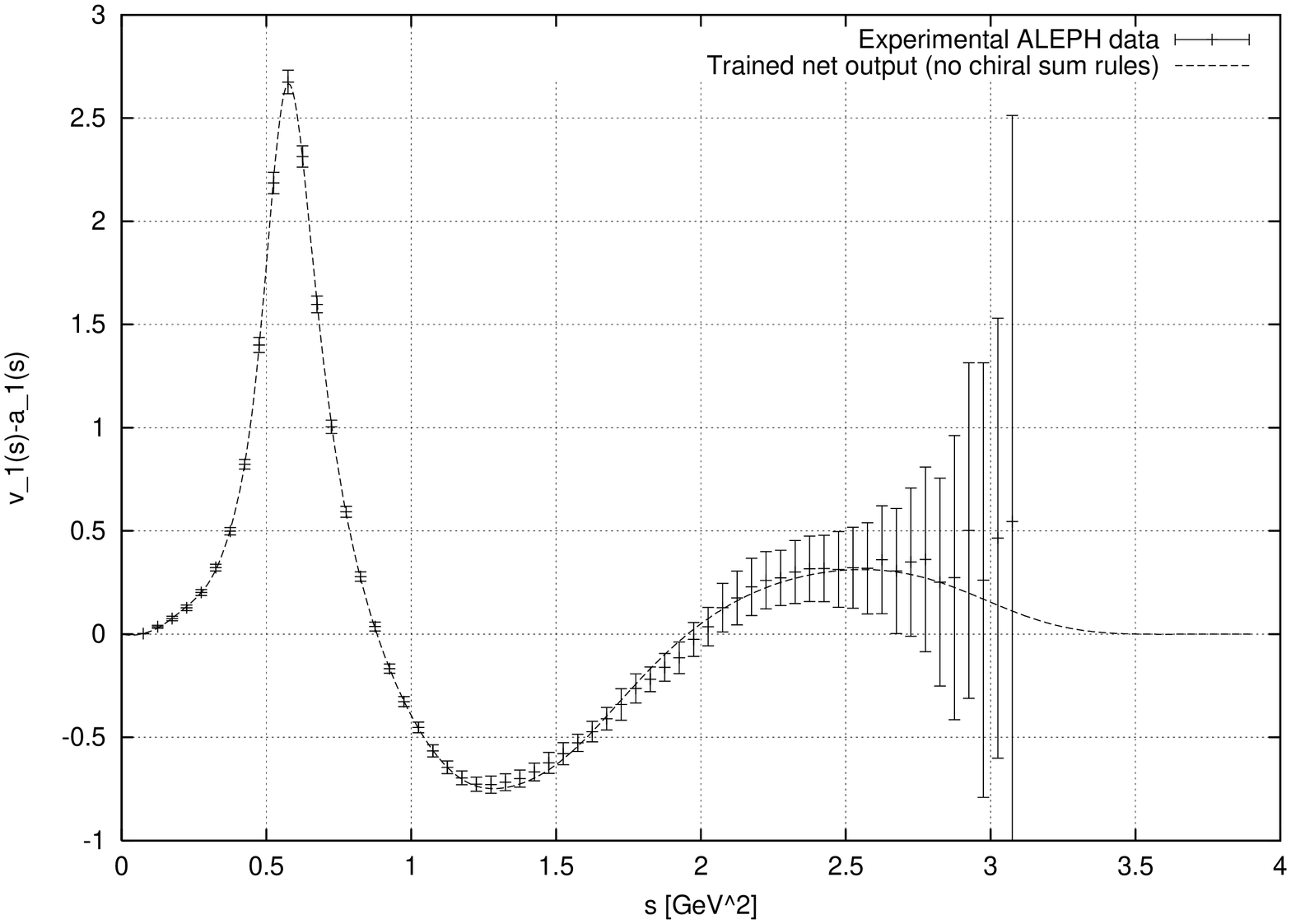}
\includegraphics[scale=0.33,angle=-90]{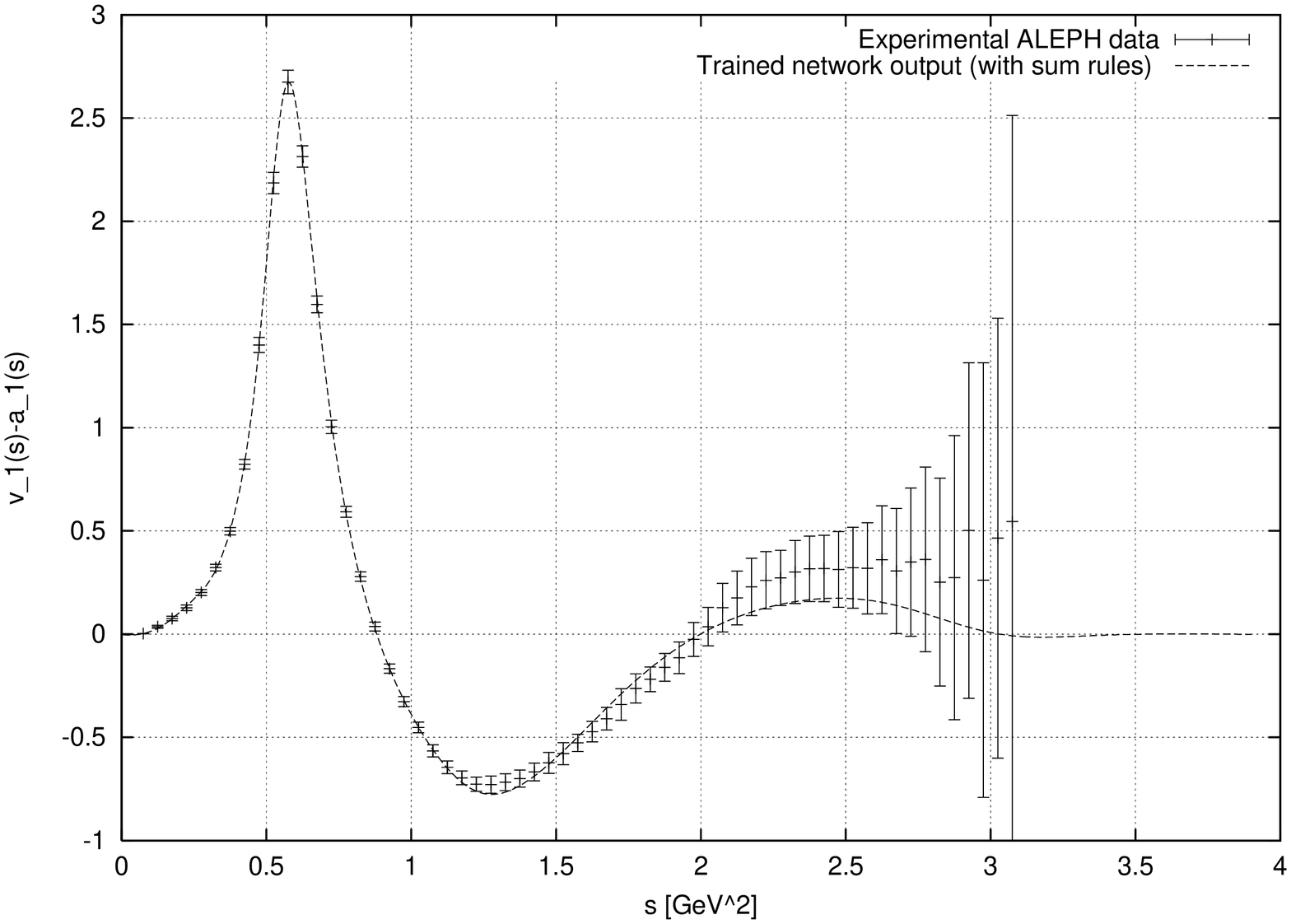}
\label{netout}
\caption{}{Output of the neural network trained over the
experimental data, without chiral sum rules (left) and
with chiral sum rules (right) incorporated in the training. Note that
the effect of the chiral sum rules is that the network output
reaches faster the asymptotic behavior of the spectral function.}
\end{center}
\end{figure}

A common problem in the genetic algorithm
 learning techniques is getting stuck in 
a local minimum of the error function, far from the absolute minimum. In
our training this difficulty has been bypassed by means of different
simple modifications of the basic training procedure. First, within 
each generation large additional mutations are performed that allow the
network configuration to
escape from local minima. Secondly, as the training advances, the 
rate of the mutations decreases, allowing for way a better 
local learning. These modifications 
are instrumental to decrease the large
duration of the training.

\subsection{Results and validation}
Once all the parameters of the training have been determined by stability
criteria, an independent set of neural networks is trained on the
spectral functions $v_1(s)-a_1(s)$.
The length of the training is fixed by studying the behavior of the 
error function $E^{(0)}$, as defined in eq. (\ref{err}) for the neural
net fitted to the central experimental values, and asking that 
$E^{(0)}/N_{dat}$ stabilizes to a value close to one, which can
be considered a good training.

A number of checks is then performed in order to make sure that an
unbiased representation of the probability density has been obtained. 
First, we have verified that the covariance of two data points
computed from the Monte Carlo sample of nets is on average very close
to the corresponding covariance matrix element of the data. Since 
correlations of the data are entirely due to systematic errors, 
this indicates that
these errors are correctly reproduced. Statistical 
estimators are then constructed as in eq. (\ref{scatterrep})  
but now referred to the trained neural networks
over the replicas, to explicitly verify that the training maintains
all the experimental information. The scatter correlation
is now defined as
\be
\label{scattertrain}
r[\rho_{V-A}^{(net)}]=\frac{\la\rho_{V-A}^{(exp)}\la
\rho_{V-A}^{(net)}\ra_{rep}
\ra_{dat}-
\la\rho_{V-A}^{(exp)}\ra_{dat}\la\la
\rho_{V-A}^{(net)}\ra_{rep}
\ra_{dat}
}{\sigma_{s}^{(net)}\sigma_{s}^{(exp)}} \ ,
\ee
and the corresponding values for the training of $N_{rep}=100$ replicas
are presented in table \ref{reptraintesttable}. 
It is seen that the central values and
the correlations are  well reproduced, whereas this is not the
case for the diagonal errors.

\begin{table}[t]  
\begin{center}  
\begin{tabular}{ccc} 
\multicolumn{3}{c}{$\rho_{V-A}(s)$}\\   
\hline
$N_{rep}$ & 10 & 100  \\
\hline  
$r [\rho_{V-A}(s)^{(net)}]$ 
  & 0.980  & 0.981  \\
\hline  
$r [\sigma^{(net)} ]$ 
  & -0.21 & -0.20  \\
\hline
$r [\rho^{(net)}]$ 
  & 0.80 & 0.85  \\
\hline
\end{tabular}
\end{center}
\caption{Comparison between experimental and generated artificial data for
  the $\rho_{V-A}(s)$ spectral function.}
\label{reptraintesttable}
\end{table}

The average standard deviation for each data point
computed from the Monte Carlo sample of nets is substantially smaller than
the experimental error. This is due to the fact that the network is
combining the information from different data points by capturing and 
underlying law, or that it is introducing a smoothing bias. 
This effect is enhanced by the inclusion of sum rules constraints. 
All networks have to fulfill these constraints which forces
the fit to behave smoothly in a region where experimental data
are very large. This should be understood as a success
of the fitting procedure.

The final set of neural networks $\rho_i^{(net)(k)}$ provides a representation
of the probability measure in the space of structure functions, which can be 
used to estimate any functional average, defined as in. eq (\ref{wa}) 
using
\be
\label{wfa}
\la\mF \lc \rho_{V-A}(s)\rc\ra=\frac{1}{N_{rep}}\sum_{i=1}^{N_{rep}}
\mF\lc \rho_{V-A}^{(net)(k)}(s)\rc.
\ee
In particular, the average and standard deviation of the
nonperturbative
condensates computed using the Monte Carlo sample will provide
a determination of the central values and errors of these
condensates.

\subsection{Details of the genetic algorithm training}
As explained above, in the second part of the
training  the chiral 
sum rules eqs. (\ref{dmo}-\ref{empionmass}) 
are incorporated to the error functional, eq. (\ref{err}).
These sum rules act as constraints on the neural network output, that is,
the main contribution to the error function (which determines the 
learning of the network) still comes from the diagonal errors, and the
sum rules are only relevant in the region where the errors are larger. 
The relative weights of the chiral sum rules will be chosen 
according to a stability 
analysis.
The effect of including sum rules in
the learning procedure  
is responsible for enforcing the desire vanishing oof
the $\rho_{V-A}(s)$ spectral function, which is
badly needed for
a reliable extraction of the nonperturbative vacuum condensates.

Obtaining  maximum stability in our output
is crucial for a proper parametrization of the spectral
function.
In the case of the relative weights of the chiral sum rules, 
we train the same network with different relative
weights and search for the region where both contributions to the error
function, the contribution from the errors and the contribution from
the sum rules, are comparable. This is shown in figure (3). 
The observed behavior is not surprising: for large relative
weights the contribution from the sum rule is small because
the training forces its learning but, as a consequence, the contribution
from the experimental data increases. This behavior can be observed explicitely
if we plot the evolution of the sum rules, as calculated with the
network output of the trained network, as a function of
their relative weights. We observe in figure (4) that, 
as the relative weight increases, the network output  better verifies
the corresponding chiral sum rule.

\begin{figure}[t]
\begin{center}
\epsfig{width=0.33\textwidth,figure=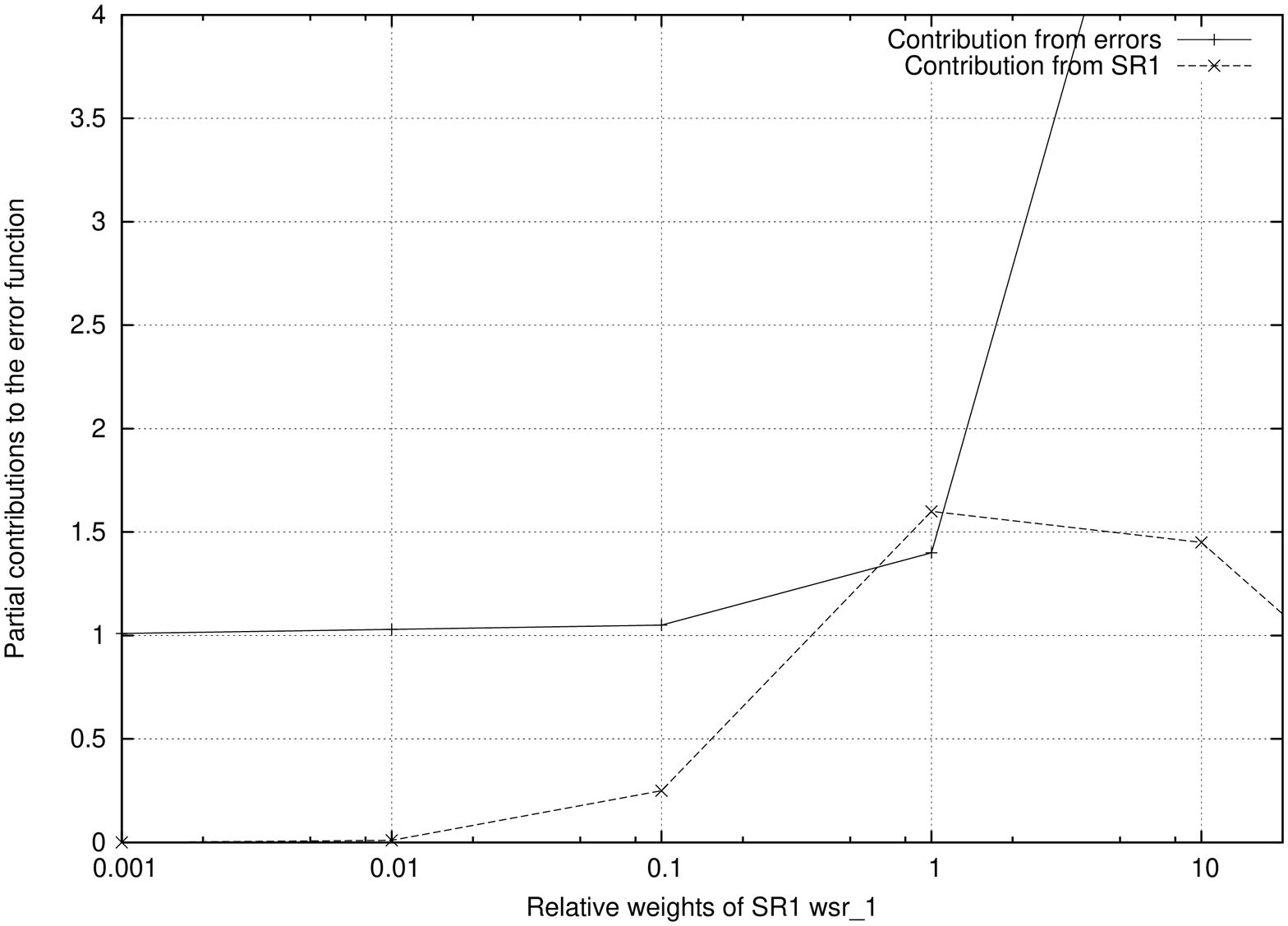,angle=-90}  
\epsfig{width=0.33\textwidth,figure=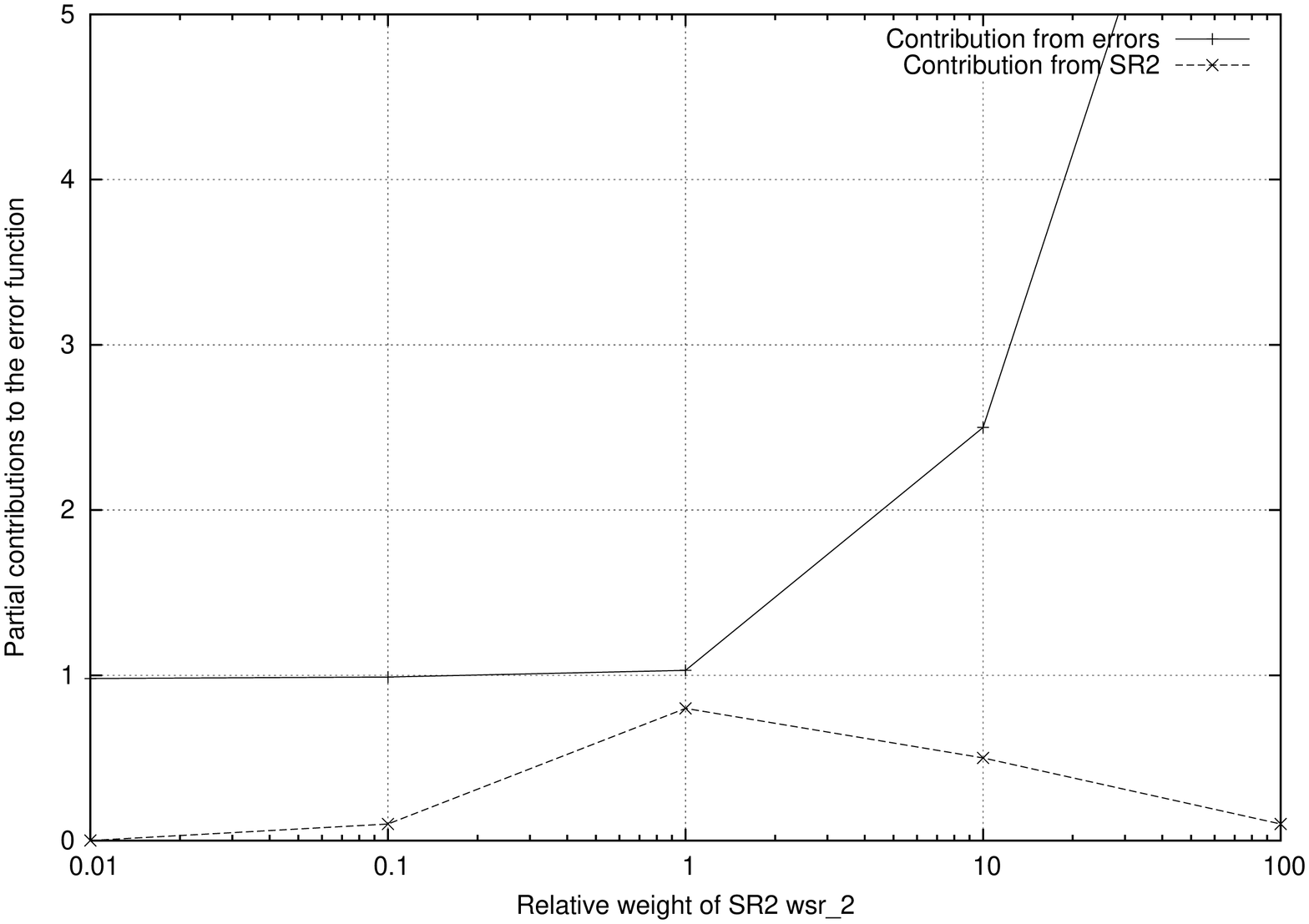,angle=-90}  
\end{center}
\caption{}{Dependence of the partial contributions to the error 
function on the relative weights of the 
SR1 (left) and SR2 (right) chiral sum rules. Note that as expected for
normalized sum rules, the stability region for the relative weight is
close to 1.
 }
\label{fitsr}
\end{figure}

\begin{figure}[t]
\begin{center}
\epsfig{width=0.33\textwidth,figure=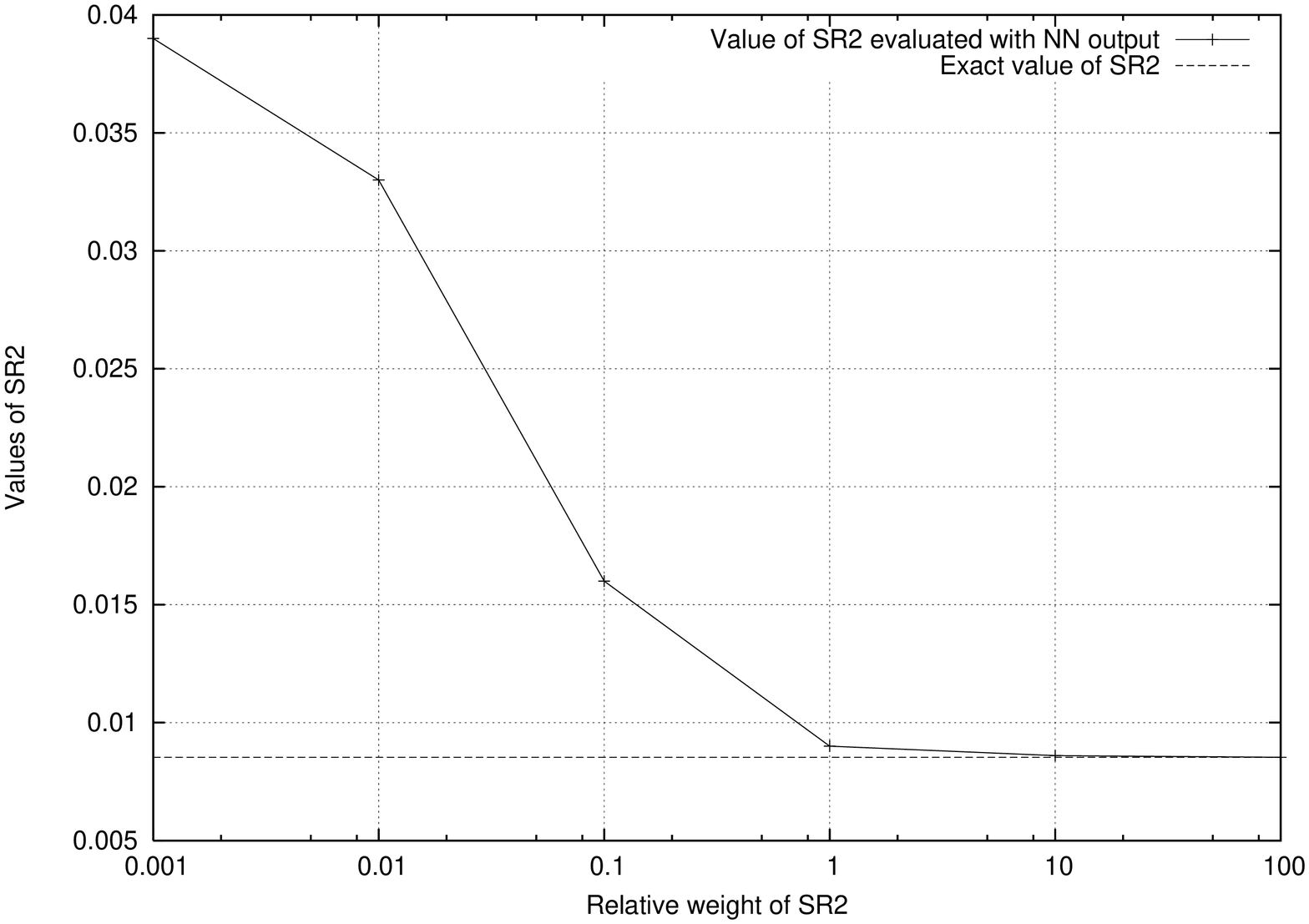,angle=-90}  
\epsfig{width=0.33\textwidth,figure=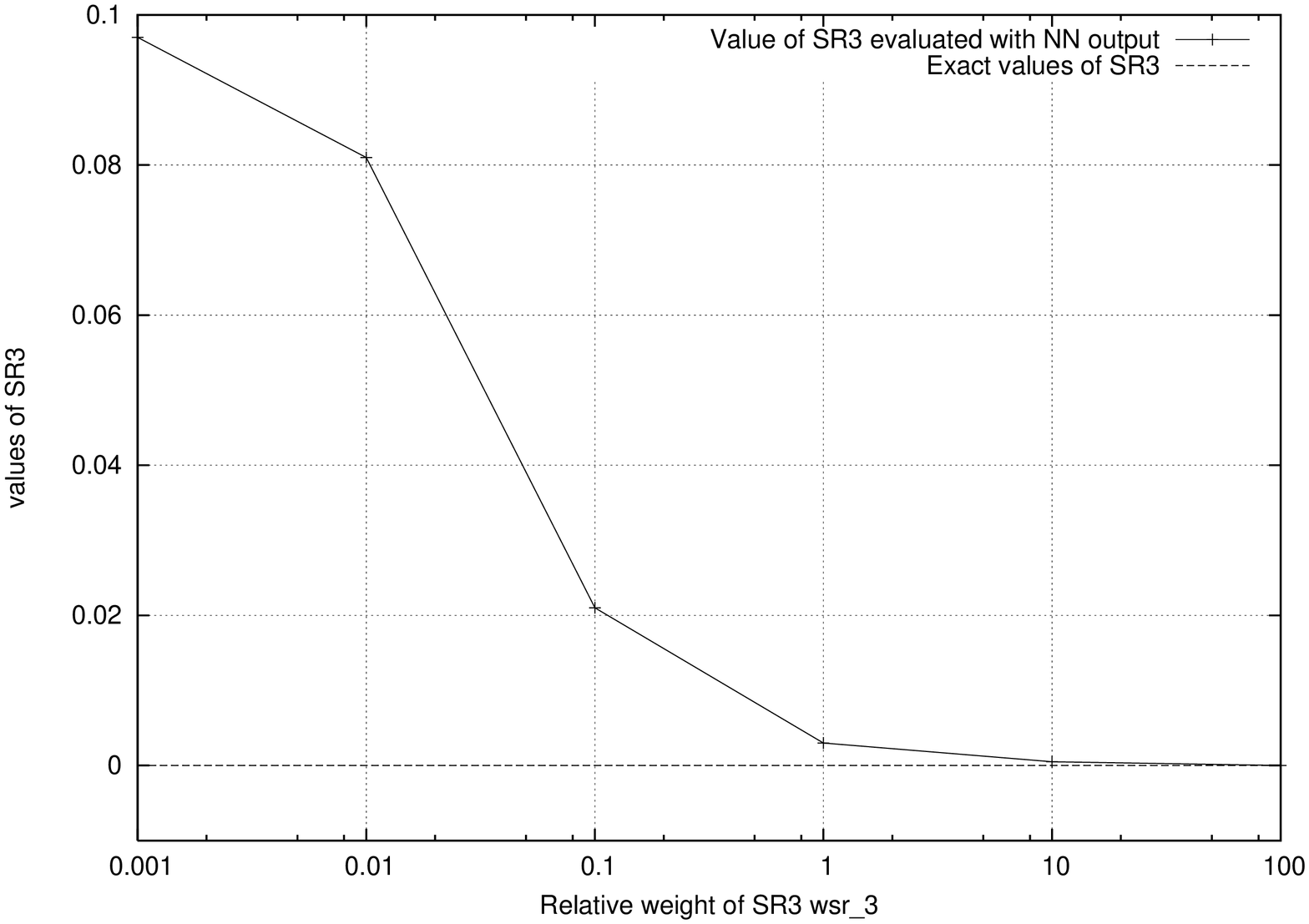,angle=-90}  
\end{center}
\caption{}{Dependence of the value of the  SR2 (left) and SR3 (right)
 chiral sum rules
 on their relative weights. It is also clear how for a value of the
relative weights close to 1 the chiral sum rules are satisfied.
}
\label{sr}
\end{figure}

\begin{figure}[t]
\begin{center}
\epsfig{width=0.45\textwidth,figure=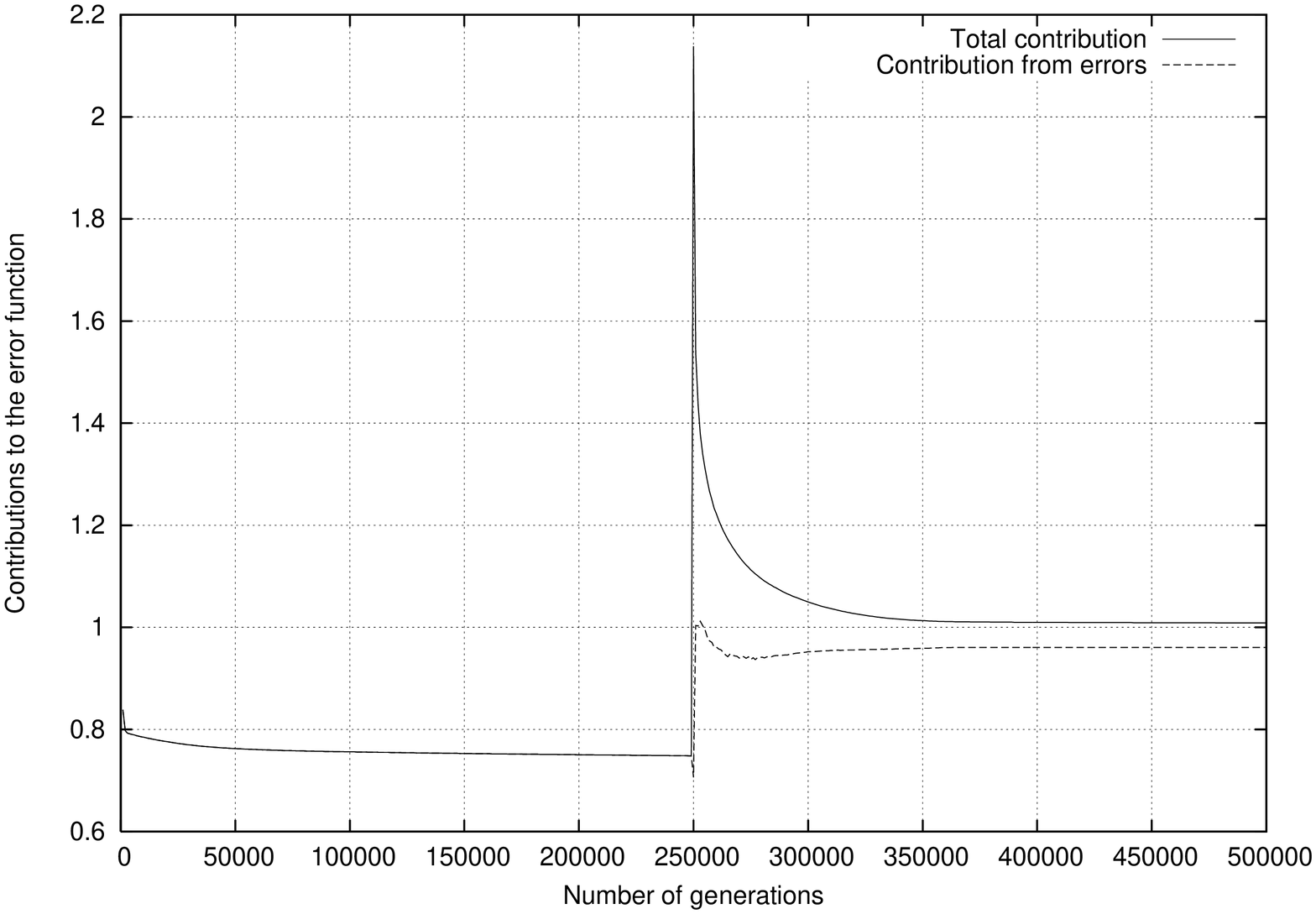,angle=-90}   
\end{center}
\caption{}{Dependence of the different contributions
to the error function on the length of the training. Note the sudden
jump when the sum rules are incorporated to the training, and how the
network later return to a configuration similar to the initial one.}
\label{evol}
\end{figure}

Genetic algorithms thus allow to implement
additional constraints in the training of neural networks in a smooth and 
efficient way. Its main drawback is the increase of the training time, 
because it is a
 random rather than a deterministic learning technique. In figure (\ref{evol})
we represent an example of the training of one replica. It can
be observed a sharp transition when the sum rules constraints are
introduced, but  
later the training forces the error function to stabilize to a 
situation similar to the initial 
training epoch. This sudden jump of the error function can be
understood as follows: when the sum rules constraints are introduced, 
the training tends to verify them, causing that the net output does
not
follow the experimental values. Nevertheless, as generations go on,
the net output begins to recover the original situation, while
maintaining the verification of the sum rules constraints.
When the number of generations is large, the error function approaches 
a value close to one, as is needed to keep systematic errors under control.

A key issue in this procedure 
is to guarantee stability of results
with respect to the relative weights of the chiral sum rules. 
In our training normalized sum rules are used, that is, if
$A_j$ is the theoretical value of the $j-$th chiral sum rule, 
the corresponding contribution to the error function will be
\be 
w_{\mathrm{sr}_j}\lp 
\int_0^{s_0}\mathrm{d}sf_j\lp\rho_{V-A}(s)\rp-A_j \rp^2{\Big /}\lp
 A_j\rp^2,
\ee
therefore we expect the relative weights
in the stability region  of order 1. The only exception is the
second Weinberg sum rule, eq. (\ref{wsr2}) 
whose relative weight has to be determined 
demanding stability of the network training, and that turns out 
to be around $w_{\mathrm{sr}_3}=10^{-1}$.

\begin{figure}[t]
\begin{center}
\epsfig{width=0.33\textwidth,figure=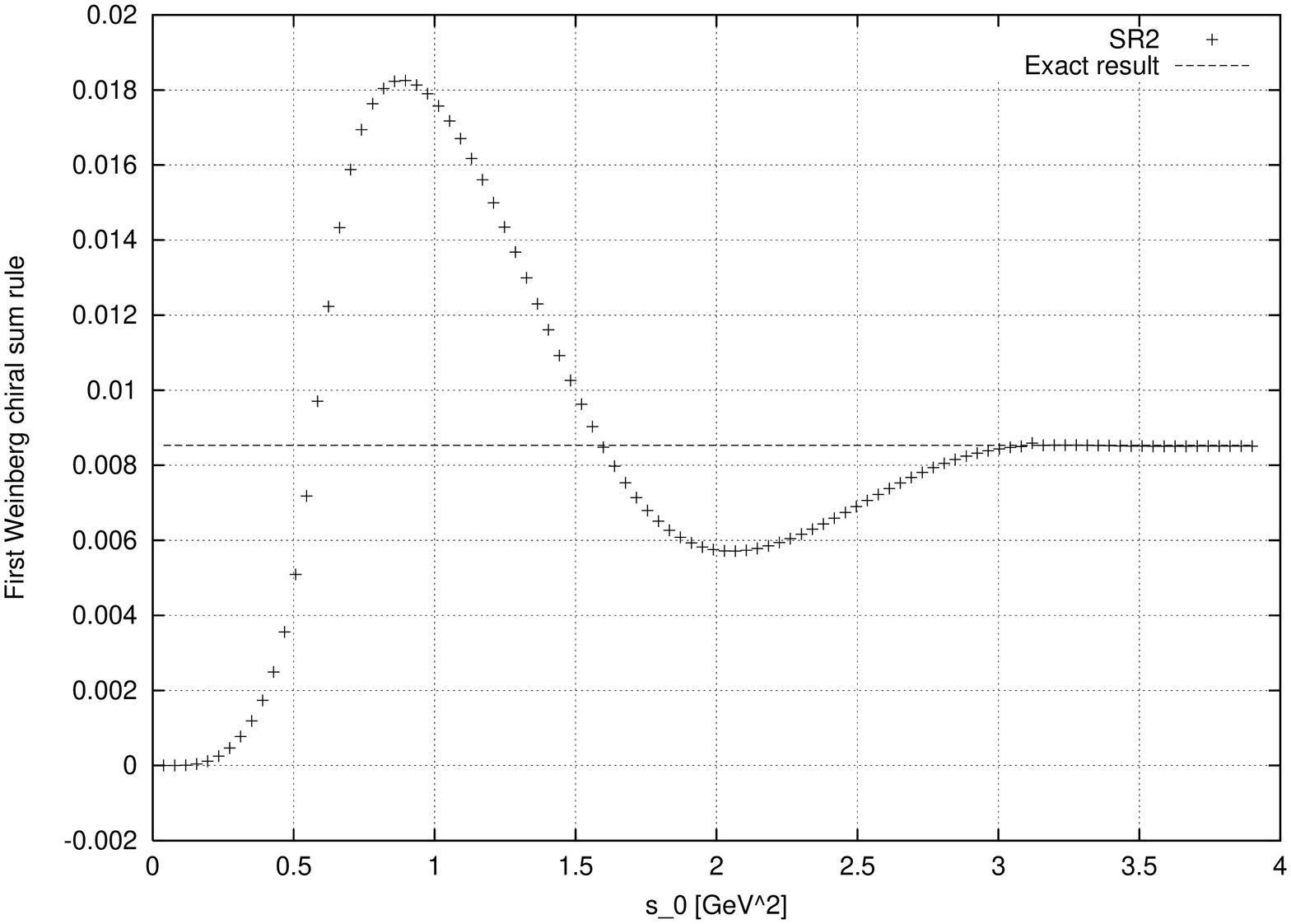,angle=-90}  
\epsfig{width=0.33\textwidth,figure=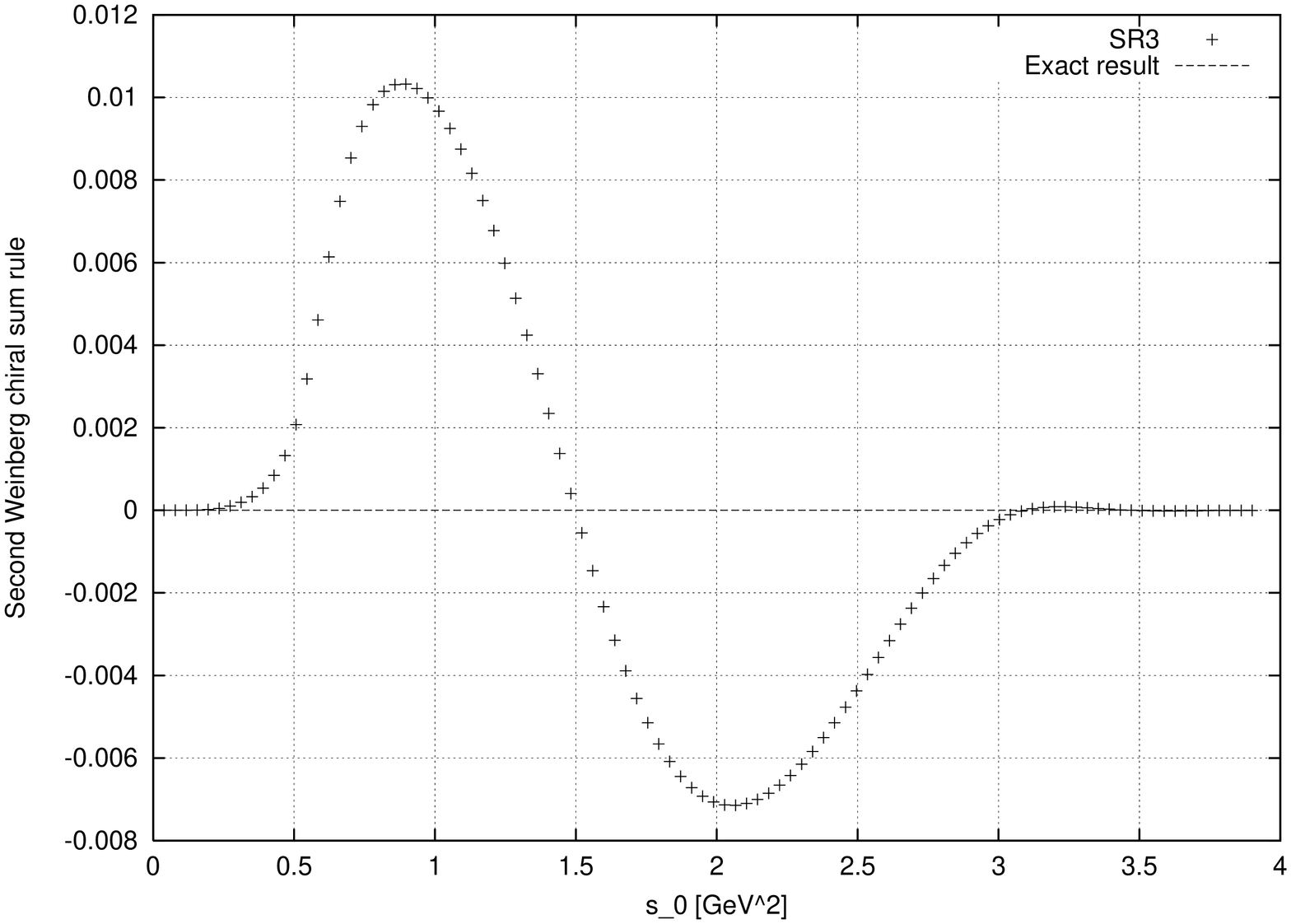,angle=-90}  
\end{center}
\caption{}{Weinberg chiral sum rules,
 SR2 (left) and SR3 (right), evaluated on the neural network
 parametrization
of $\rho_{V-A}(s)$}
\label{srstab}
\end{figure}

Let us emphasize that the two Weinberg chiral sum rules are
well
verified by our neural network parametrization, and thus have been
incorporated to the information contained on the 
experimental data. This fact will be  crucial later because 
different extraction methods, differing in combinations
of these chiral sum rules, can be shown to be equivalent in the 
asymptotic region $s_0\to\infty$. In figure (\ref{srstab}) the 
two Weinberg sum rules, eqs. (\ref{wsr1},\ref{wsr2}) evaluated with
the neural network parametrization of the spectral function 
$\rho_{V-A}(s)$ are represented. Both chiral sum rules are well
verified
in the asymptotic region, beyond the range of available 
experimental data. This also will ensure the stability of the
evaluation of the condensates with respect to the
specific value of $s_0$ chosen as long as it stays in the
asymptotic region.

\section{Determination of the nonperturbative condensates}
Using the neural parametrization of spectral functions, we can compute
for each trained replica any given sum rule. Because the neural 
parametrization retains all the experimental information (it 
even allows for a determination of errors and correlations), we can view
values coming from the neural networks
 as direct experimental determinations of convolutions
of the spectral function $\rho_{V-A}(s)$.
The value of the  condensates $\la\mo_6\ra,\la\mo_8\ra$  and higher 
dimensional condensates is then extracted
from the value of an appropriate sum rule, eq. (\ref{condenconv}).
The method we will follow is the evaluation of the vacuum condensates
as a function of the upper limit of integration for each replica
and compute the mean and standard deviation. As has been
explained before, it is crucial to represent the value of the
different sum rules as a function of the upper limit 
of integration, to check both its convergence and its stability. 

Our method works as follows. First, we train a neural network on each replica.
On a first training epoch we do not use the sum rules, so that the training
can arrive to the best possible minimum. This is important when
training neural networks because when further constraints to the training
are added, as in our case the chiral sum rules, the goodness
of the fit will be better if we start from a deep local minimum.
On a second training epoch,
we add to the fitness the contribution from the sum rules, where the
relative weights are chosen so that the sum rules never represent more
than the contribution from the experimental errors to the total fitness. 
Then, sum rules
act as a smooth constraint on the network training, being more
relevant in the regions with larger errors and thus enforcing
the asymptotic vanishing of the spectral function. 
This technique prevents the contribution
of the chiral sum rules to become 
so strong that overcome the experimental data with the corresponding errors.

\subsection{Central values}
The first criterion to judge the reliability of a QCD sum rule
is its independence, at large values of $s$ from the value of the upper
integration limit, that its, its saturation. We then need
to explore the values for the final condensates which are
stable against the limit of integration of the sum rule. This
stability criterium is completed with demanding independence
of the results on the specific polynomial entering the sum rule.
Further criteria are stability with respect to the precise
artificial endpoints added to the data and with respect to
the relative weights in the error function used to train
the neural networks.  

Stable results are obtained for the dimension six condensate $\la \mo_6 \ra$ 
whereas higher condensates {\it e. g.}  $\la \mo_8 \ra$ are less stable. 
Fig. \ref{condenextracterr}
shows the outcome for $\la \mo_6\ra$ and
$\la \mo_8 \ra$ including the propagation of
statistical errors. It is clearly seen that convergence
in the limit of integration $s_0$ is obtained
due to the addition of sum rules and endpoints in the
learning procedure. The central values for the
condensates in the asymptotic limit come out to be:
$s\to\infty$:
\begin{eqnarray}
 \la \mo_6 \ra= -4.2~10^{-3}~\mathrm{GeV}^6 \ ,
\nonumber
\\
\la \mo_8 \ra= -12.7~10^{-3}~\mathrm{GeV}^8 \ .
\end{eqnarray}
The
value of the $\la \mo_6 \ra$ is a cross check of the validity of our 
treatment: not only there are strong theoretical arguments that
support
the fact that $ \la \mo_6 \ra$ is negative
\cite{witten},\cite{latorre} 
but also all previous determinations with different techniques yield
negative results, being the majority of them compatible with
ours within errors.

We note that our evaluation of the condensates is compatible
with some of our previous evaluations and has a similar error.
This is though misleading as the error quoted here is only
statistical and a discussion on systematic errors is needed
(and done below). 
We can also obtain  values for the higher dimensional
nonperturbative condensates:
\bea
 \la \mo_{10} \ra= 7.8~10^{-2}~\mathrm{GeV}^{10} \ ,\nonumber \\
 \la \mo_{12} \ra=-2.6~10^{-1}~\mathrm{GeV}^{12} \ .
\eea
Although  stability deteriorates as compared to the case of the lower
dimensional condensates,
these central values for the condensates are alternated in sign.

\subsection{Discussion of errors}
The discussion of the
various
sources of errors is  crucial  to our treatment.
We enumerate and discuss then in turn:

\begin{enumerate} 
\item Statistical error propagation from the experimental covariance
  matrix.\\
This is the best understood and treated error source
in our analysis. As explained above, the neural network
  parametrization defines an unbiased
probability measure in the space of spectral functions
 that provides a nonlinear
error propagation. This source of error is  kept under
control by using the averages over Monte Carlo replica.
The band for central values of the condensates 
allowed by this error propagation can be
visualized in  fig. (7). 
Numerically, the contribution to the experimental error 
(statistics and correlations) to the central values is
\begin{eqnarray}
 \la \mo_6 \ra= \lp-4.2\pm 1.1\rp ~10^{-3}~\mathrm{GeV}^6 \ ,
\nonumber
\\
\la \mo_8 \ra= \lp -12.7\pm 6.4\rp~10^{-3}~\mathrm{GeV}^8 \ ,
\nonumber
\\
 \la \mo_{10} \ra= \lp 7.8\pm 2.4 \rp~10^{-2}~\mathrm{GeV}^{10} \ ,
\nonumber
\\
\la \mo_{12} \ra= \lp -2.6 \pm 0.8 \rp~10^{-1}~\mathrm{GeV}^{12} \ .
\label{staterr}
\eea
Note that the sign obtained for each condensate remains unaltered
within error.

\begin{figure}
\begin{center}
\label{condenextracterr}
\includegraphics[angle=-90,scale=0.28]{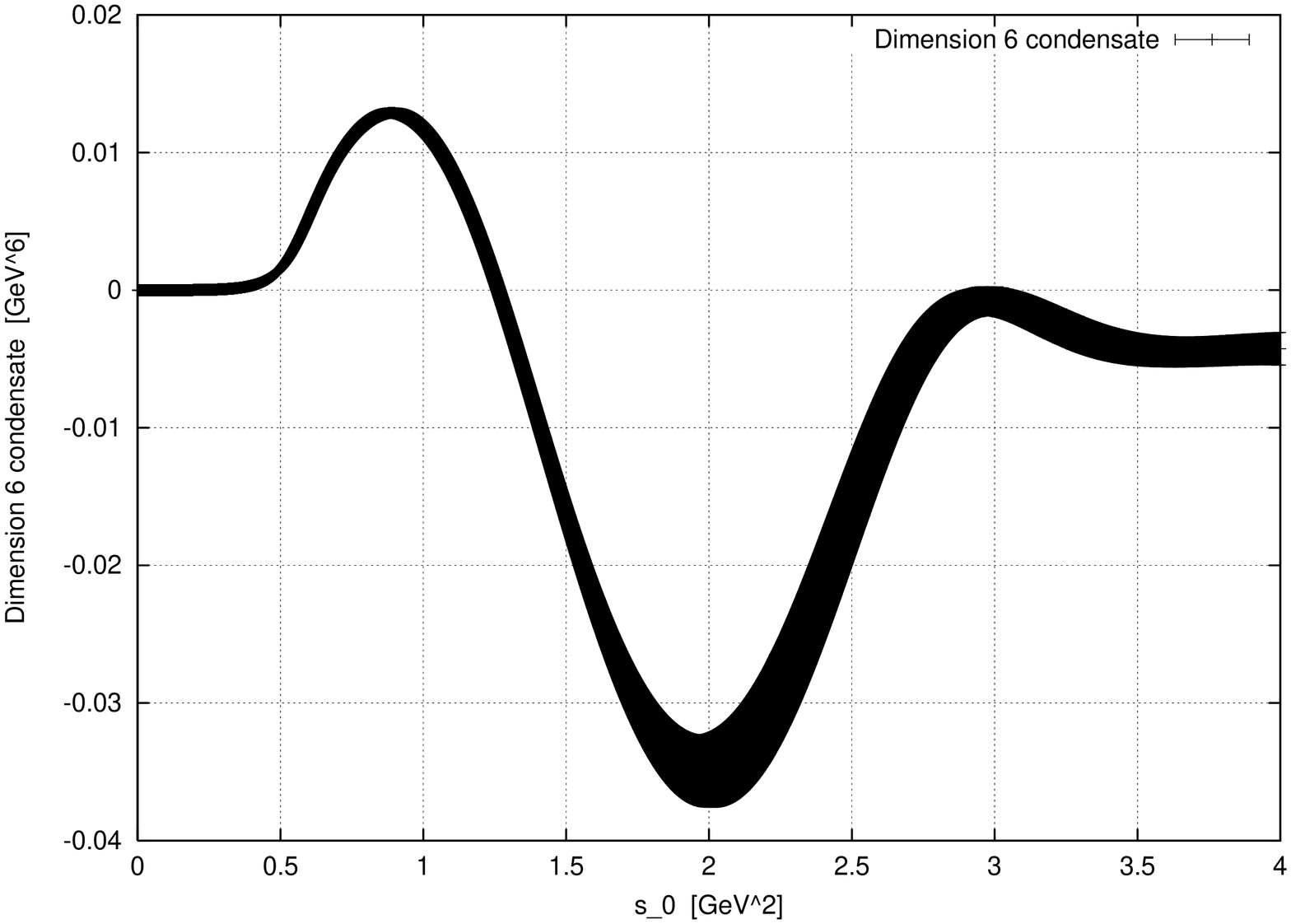}
\includegraphics[angle=-90,scale=0.28]{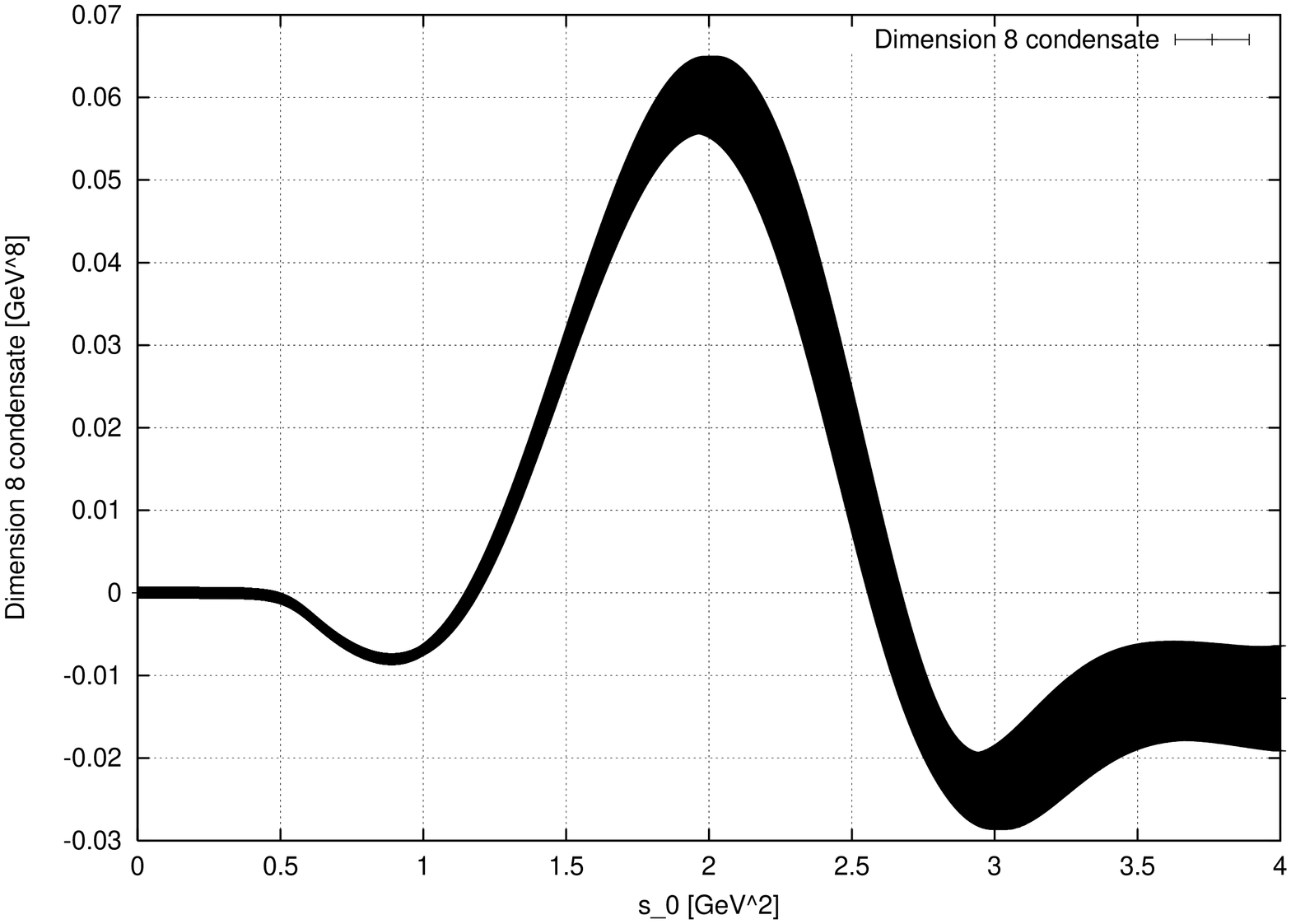}
\caption{}{Condensates $\la \mo_6 \ra$ and  $\la \mo_8 \ra$ as a function of
  $s_0$. The error bands only include  the propagation of experimental
uncertainties.}
\end{center}
\end{figure}

\item Choice of the polynomial in the finite energy sum rule.\\
In principle, there are potentially important
systematic uncertainities coming from the
method of extraction of the condensates. These are much more
difficult to assess, as noted when looking through the extense available
literature. The extraction of 
$\la \mo_6 \ra$ turns out to be  clean and its errors are
essentially of statistical nature.
The uncertainty increases with the dimension of the condensate. 
Let us elaborate further these statements.

Consider the 
following convolutions
\be
Z_{6a}\equiv \int_0^{s_0}ds\frac{1}{2\pi}s^2\rho_{V-A}(s)
\ee
\be
Z_{6b}\equiv s_0^2\int_0^{s_0}ds\frac{1}{2\pi}\lp1-\frac{s}{s_0}\rp^2
\rho_{V-A}(s)-f_{\pi}^2s_0^2
\ee
The second equation is only equivalent to the first if, for some
$s_0$,
both Weinberg sum rules are satisfied. Although 
experimental data on tau decays do not exactly saturate these sum rules, 
the neural network parametrization trained to
obey all the sum rules  showed that Weinberg sum rules
can indeed be well verified in the asymptotic region. 
We should then expect that
$Z_{6a}$ and
\be
Z_{6b}\equiv \int_0^{s_0}ds\frac{1}{2\pi}\lp s^2-2s s_0\rp
\rho_{V-A}(s)
\ee
yields similar results for $\la \mo_6 \ra$ within errors, as can be
seen in fig. (\ref{srcond}). The same applies to the dimension
8 condensate $\la \mo_8 \ra$, where we now define the following
finite energy sum rules: 
\be
Z_{8a}\equiv -\int_0^{s_0}ds\frac{1}{2\pi}s^3\rho_{V-A}(s) \ ,
\ee
\be
Z_{8b}\equiv -\int_0^{s_0}ds\frac{1}{2\pi}\lp s^3-ss_0^2\rp 
\rho_{V-A}(s) \ ,
\ee
\be
Z_{8c}\equiv -\int_0^{s_0}ds\frac{1}{2\pi}\lp s^3+3s_0^2s\rp
\rho_{V-A}(s) \ .
\ee
We conclude that the neural network parametrization of 
spectral functions properly trained to accomodate for all sum
rules provides estimates for condensates which are 
independent of the choice of a specific finite energy sum rule.

\begin{figure}[t]
\begin{center}
\epsfig{width=0.33\textwidth,figure=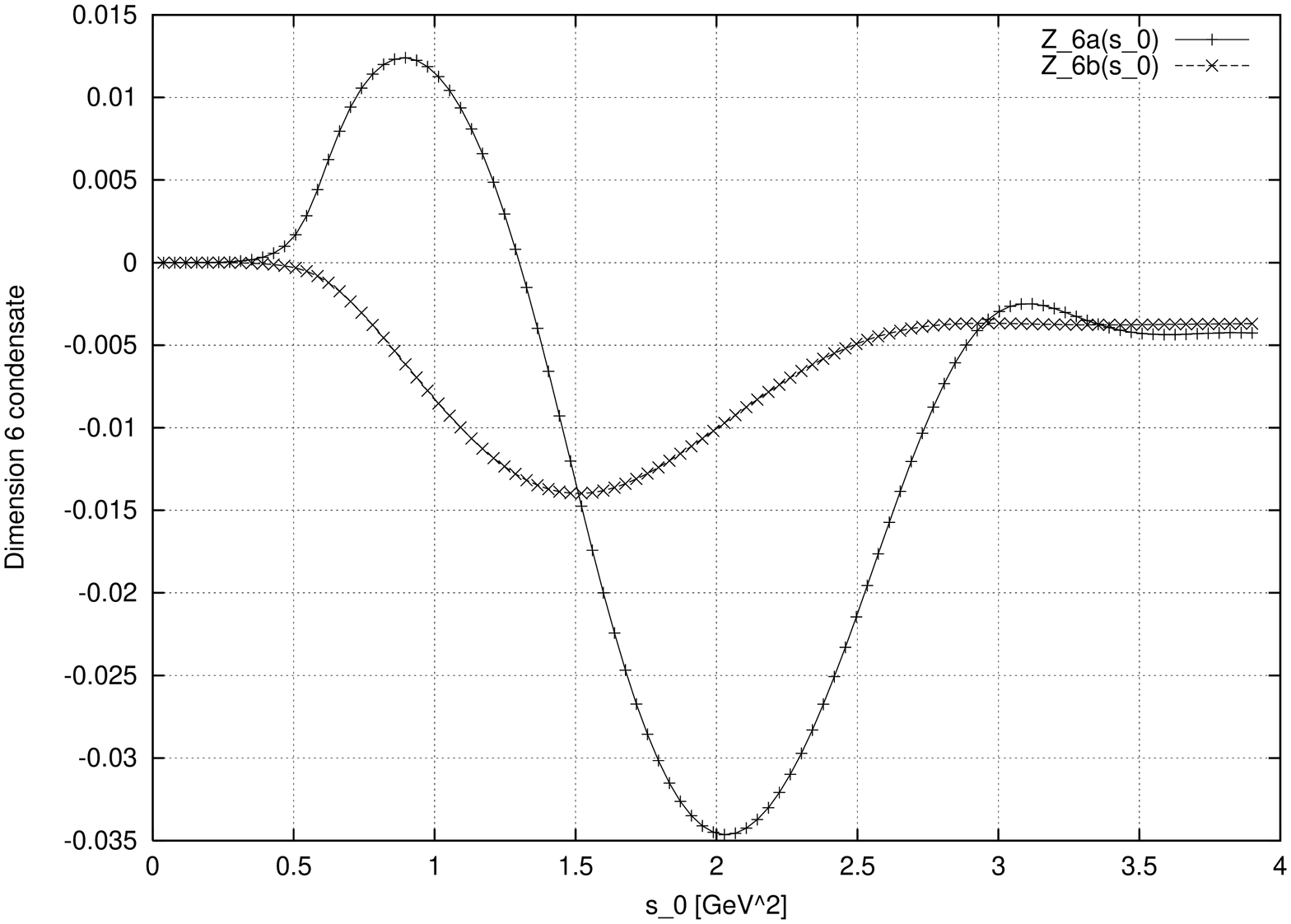,angle=-90} 
\epsfig{width=0.33\textwidth,figure=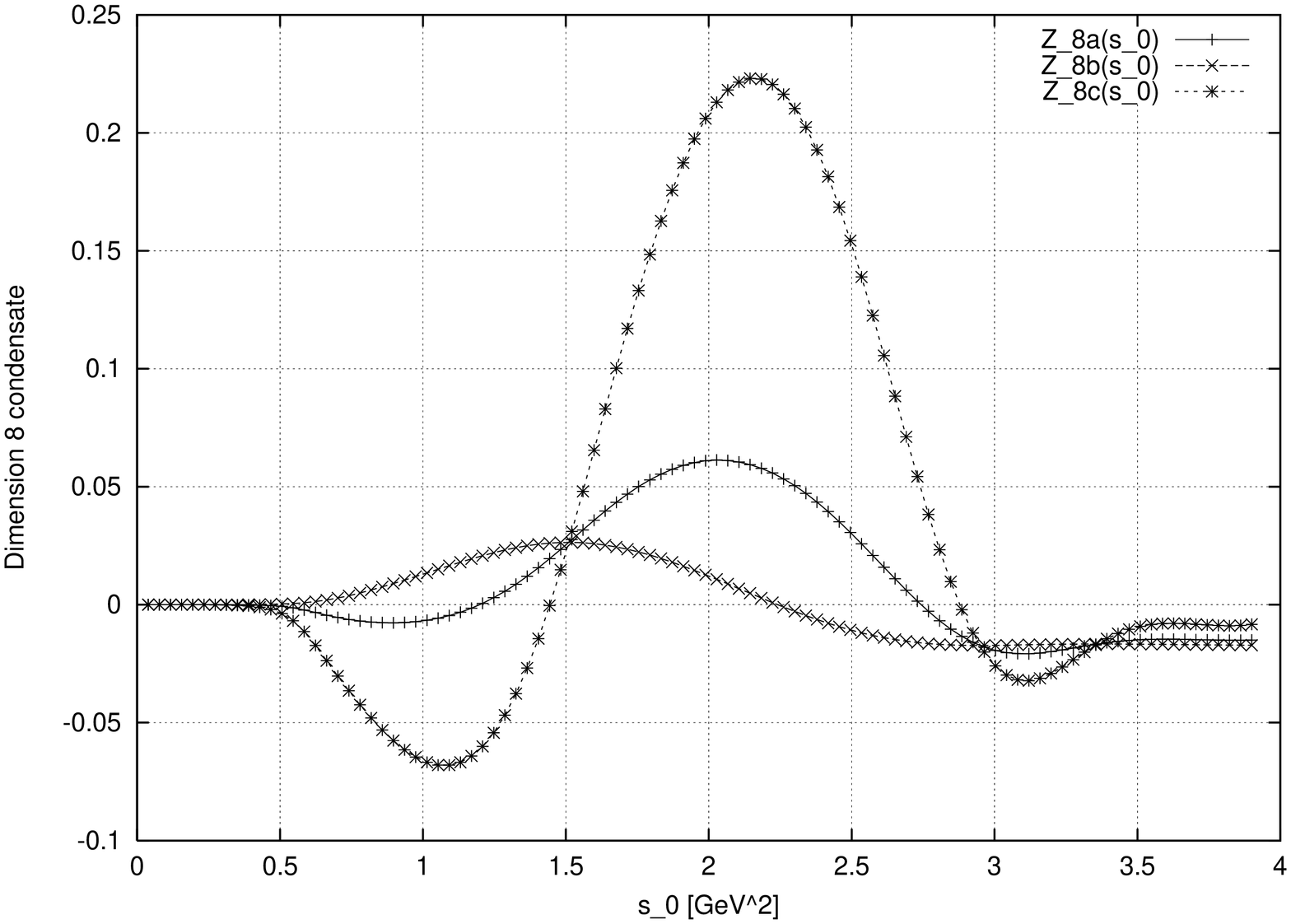,angle=-90}    
\end{center}
\caption{}{Extraction of $\la \mo_6 \ra$ and $\la \mo_8 \ra$ 
 with different polynomials}
\label{srcond}
\end{figure}

\item Dependence on the endpoints.\\
Our fit
implements the asymptotic constraint that $\rho_{V-A}(s\to\infty)=0$
by adding artificial endpoints. It is then necessary to
verify the degree of sensitivity of our ouput to the precise location
of these endpoints. As shown if Fig.  (\ref{stabend}), the
sign of the dimension six and eight condensates 
remains unaltered when endpoints range  between 3.5 and 4 GeV$^2$.
We also
observe relatively large, although compatible within errors, 
fluctuations of the central values. This effect
may be related   to the presence
of small wiggles in the spectral function $\rho_{V-A}(s)$ 
for large $s$.  The contribution
of this source of uncertainty to $\la \mo_8 \ra$ turns to
dominate over statistical uncertainties and can be estimated 
to be
\be
-2.5~10^{-2} \le \la \mo_8 \ra \le -5~10^{-3}  
\ee
while for $\la \mo_6 \ra$ is comparable to the uncertainty due
to the statistical errors of the experimental data, that is,
\be
-6~10^{-3} \le \la \mo_6 \ra \le -2~10^{-3}  \ .
\ee

\begin{figure}[t]
\begin{center}
\epsfig{width=0.33\textwidth,figure=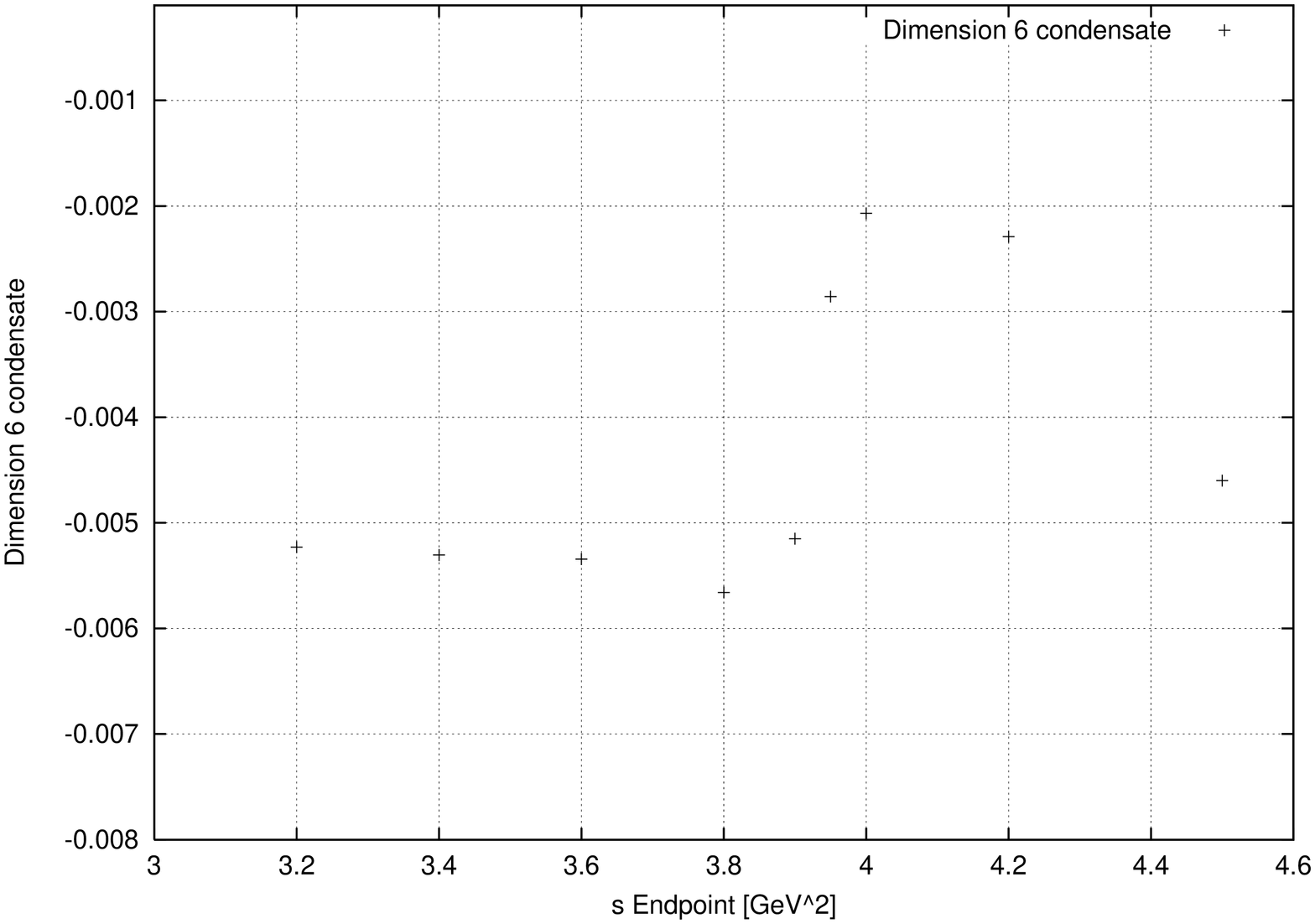,angle=-90} 
\epsfig{width=0.33\textwidth,figure=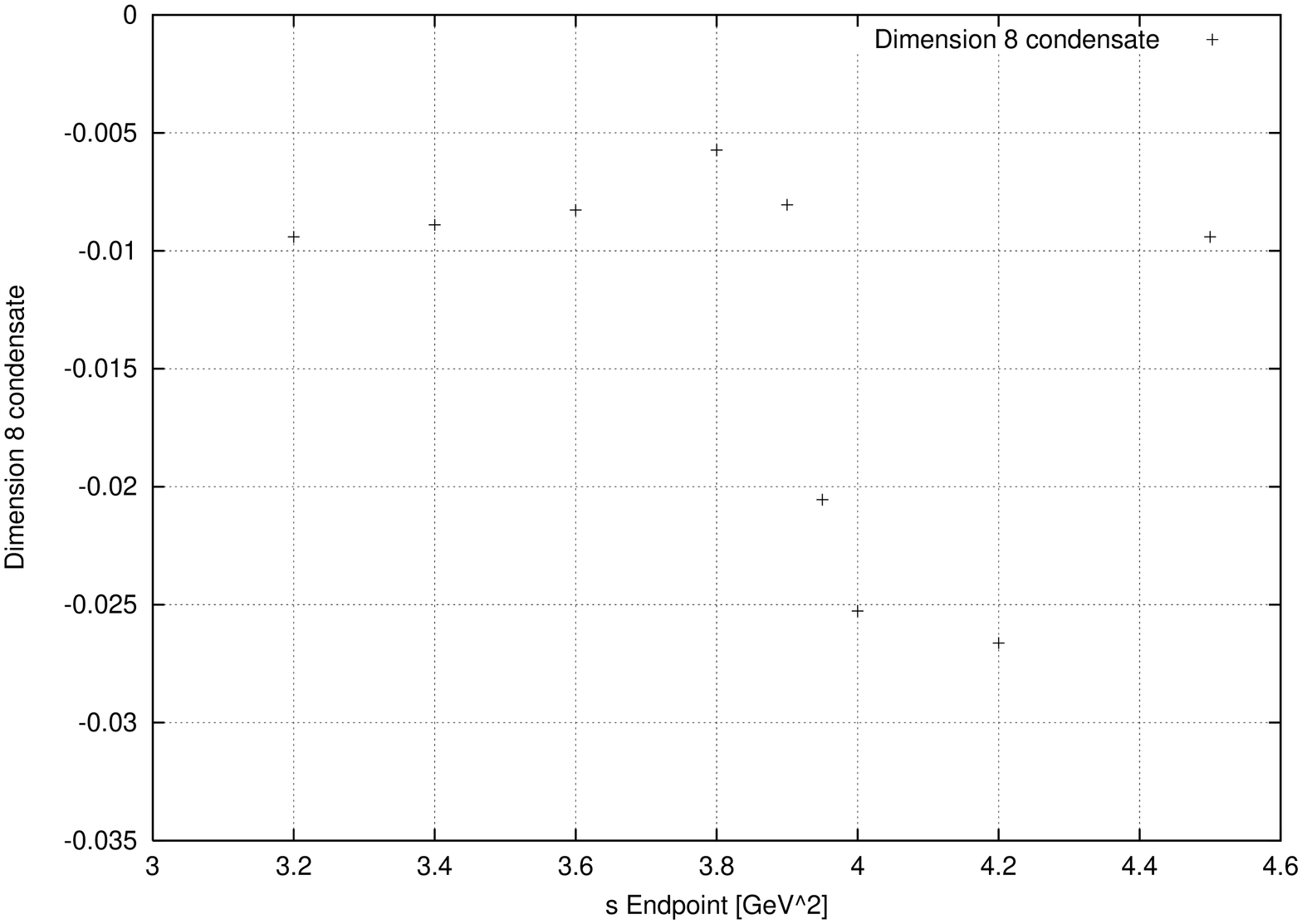,angle=-90}    
\end{center}
\caption{}{Value of the condensates $\la \mo_6 \ra$ (left) and 
$\la \mo_8 \ra$ (right) as a function
of the position $s$ of the first artificial endpoint. Note that their sign
remains unchanged and the presence of a stability region near $s=3.5$ GeV$^2$}
\label{stabend}
\end{figure}

\item Chiral sum rules. \\
This turns out to be the main source of systematic errors for
the dimension 6 condensate.
Chiral sum rules can be forced to be fulfilled by adequate
training to any desired degree of precision. This, though,
introduces a large increase 
in the total error function, eq. (\ref{err2}), coming
from the experimental error piece. It is then
necessary to make an appropriate choice of relative weights
between the error associated to experimental data and
the error associated to the fulfillment of chiral sum
rules.

We may advocate that 
the most appropiate relative weights for the  normalized
chiral sum rules are $\mo(1)$. This is due to the
fact that the total error function jumps above 1 
when too large relative weights are considered, as seen in 
figs. (\ref{fitsr}, \ref{sr}). We have thus performed
a multi-dimensional stability analysis searching for the
relative weights that produce a minimum sensitive
final result, supplemented with the condition
that in any case the contribution from the experimental errors to the
error function can be greater than 1. 
The most suitable relative weights for the chiral sum rules 
turn out to be 
\be
w_{sr1}=1.0 \quad w_{sr2}=5~10^2 \quad
w_{sr3}=0.3 \quad w_{sr4}=1~10^2 \ . 
\ee
This stability analysis 
shows that the sign of the central values  of the
condensates is not very sensitive to the relative weights
for the chiral sum rules. The estimation of the
error associated with this uncertainty leads to the following
range of values for the lowest dimensional condensate
\bea
 -2~10^{-3} \le\la \mo_6 \ra \le -6~10^{-3} ~\mathrm{GeV}^6 \ . 
\eea
For $\la \mo_8 \ra$ the statistical errors and the systematic error
due to the position of the artificial endpoints turn out to dominate 
over this source of uncertainty.
Similar estimates for the condensates of higher dimensions turn out
not to be reliable, and therefore we present only the central values obtained 
in this analysis together with the statistical errors.

\end{enumerate}

\subsection{Analysis of the $s_0=1.5$ GeV$^2$ {\it duality point}}
Some values of previous extraction of the
condensates
\cite{periscond} are based on the existence of a {\it duality point}
around $s_0\sim 1.5$ GeV$^2$. 
Our neural network parametrization is such that
 the second Weinberg sum rules is indeed verified 
around this point. Consequently,
 the values of the condensates computed by truncating 
different finite energy sum rules at this point do agree,      
 as can be seen in fig. (\ref{srstab}). 
Nevertheless, as shown in fig. (\ref{srcond}), the value of
the condensates at this duality point is different than the 
asymptotic one. Both results share the same sign but not the
same absolute
magnitude for $\la\mo_6\ra$, and with the opposite sign for
$\la\mo_8\ra$ (which is precisely the results that the authors in 
Ref. \cite{periscond} obtain).

These extractions using the first duality point can be justified
by some resonance models of the hadronic spectrum \cite{peris}, inspired in the
large$-N_C$ limit of QCD with the additinal assumption
of the validity of the Minimal Hadronic Approximation (MHA), 
in which the spectral functions are saturated by the pion pole, the
first axial vector and the first rhoo vector resonances.
Experimentally, it is observed that at
different {\it duality points}, even as low as $1.5$ GeV$^2$, there
is a local quark-hadron duality, meaning that the OPE at the quark-
gluon level and that evaluated with the entire resonance hadronic
spectrum coincide. Whether this apparent {\it duality point} 
at  $s_0=1.5$ GeV$^2$ is an accident only due to the
fullfillment of the second Weinberg sum rule, or it is really a 
consequence of the full QCD hadronic spectrum remains to be
understood. What it is clear from our analysis is that
the condensates evaluated at this first duality point are 
different to those obtained in the asymptotic regime $s_0\to\infty$,
where the validity of the OPE is less questioned.

\subsection{Spectral functions and the electroweak penguin $Q_7$}
As a byproduct of our analysis additional sum rules of the
spectral function $\rho_{V-A}$ which are relevant to phenomenology 
can be estimated. As an example\footnote{In this section the
work of Ref. \cite{prades} is followed, the reader is directed to this
reference for definitions and notation}, the 
sum rule 
\be
\label{alr}
\mathcal{A}_{LR}(\mu^2)\equiv\int_0^{s_0}ds~s^2\ln\lp\frac{s}{\mu^2}\rp
\frac{1}{2\pi^2}\rho_{V-A}(s) \ ,
\ee
will be considered,
where $\mu^2$ is a arbitrary factorization scale that cancels in the 
computation of physical observables. In this analysis the value
$\mu^2=2$ GeV$^2$ will be used. Eq. (\ref{alr}) is relevant to
the evaluation of $\mathrm{Im}~G_E$, where $G_E$ is one of the 
couplings of the low energy chiral Lagrangian describing $|\Delta
S|=1$
transitions \cite{prades}. The importance of a precise determination of this
coupling
relies on the fact that $\mathrm{Im}~G_E$ is one of the most important
contributions to $\epsilon'$ in the Standard Model. In turn, its value
is
dominated by the electroweak penguin contributions $Q_7$ and $Q_8$,
which
explains why the data on spectral functions from hadronic tau decays
is important in its determination.

\begin{figure}
\begin{center}
\label{alrfigure}
\includegraphics[angle=-90,scale=0.4]{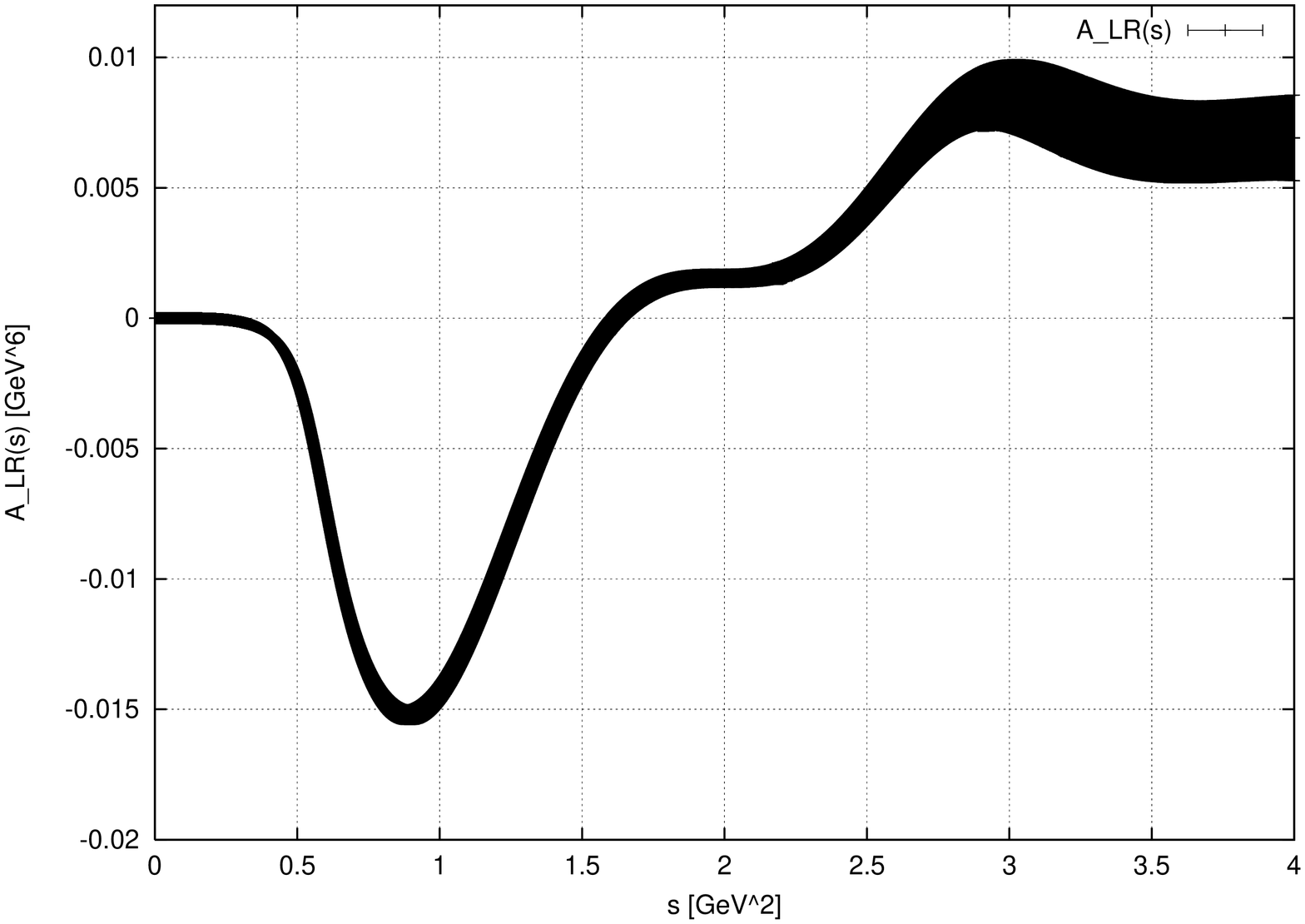}
\caption{}{Evaluation of eq. (\ref{alr}) for diferent values of
  $s_0$. 
The error bands include the propagation of experimental
uncertainties.}
\end{center}
\end{figure}

Following the same steps that lead to the determination of the
vacuum condensates, the same procedure for the sum rule
eq. (\ref{alr})
is repeated. The result that is obtained 
in the asymptotic limit $s_0\to\infty$ is
\be
\mathcal{A}_{LR}=\lp 6.9 \pm 1.6\rp ~10^{-3}~\mathrm{GeV}^6 \ ,
\ee
as can be seen in figure (10). It should be noted that
the present determination is in good agreement with that
obtained in the original work\footnote{Note however that
a different normalization for the spectral correator is used.}
\cite{prades}. The quoted error only refers to the propagation
of experimental uncertainties.

\subsection{Results and comparison with other determinations}
Our  determination of the nonperturbative condensates
including the statistical error coming from the experimental
errors and correlations was given in eq. (\ref{staterr}).
However,  the error which dominates the determination of 
$\la\mo_6\ra$ 
comes from the relative weights of the chiral
sum rules to be obeyed.
We have performed a stability analysis on these
relative weights that produces
a final result: 
\be
 \la \mo_6 \ra=\lp -4.0 ~\pm 2.0\rp~10^{-3}~\mathrm{GeV}^6 \ .
\nonumber
\ee
As explained above, for the dimension 8 condensate, the systematic
error
associated with the endpoint position is comparable to the
statistical uncertainty, that combine to yield a value
\be
\la \mo_8 \ra=\lp -12 ~\epm{7}{11} \rp 10^{-3}~\mathrm{GeV}^8 \ . 
\ee
For higher dimensional condensates it is much difficult to estimate the
different sources of systematic uncertainties. We, then, quote
the central values we obtained and their statistical error:
\bea
 \la \mo_{10} \ra= \lp 7.8\pm 2.4\rp~10^{-2}~\mathrm{GeV}^{10} \ ,
\nonumber
\\
\la \mo_{12} \ra=\lp-2.6\pm 0.8 \rp~10^{-1}~\mathrm{GeV}^{12} \ .
\label{finalresults}
\eea
A similar analysis has been performed with
the OPAL data, yielding equivalent results but with larger errors, due
to the larger statistical uncertainties as
compared to the ALEPH experimental
data. The values of the 
QCD nonperturbative condensates have been previously 
extracted from the ALEPH and OPAL data, with different techniques and
different results, as summarized in table \ref{lit}.

\begin{table}[t]  
\begin{center}  
\begin{tabular}{ccc}  
\hline
\hline
 Reference & $ \la \mo_6 \ra \times 10^{3} \mathrm{GeV}^6  $ 
&  $ \la \mo_8 \ra \times 10^{3} \mathrm{GeV}^8  $  \\
\hline  
 
 Ref. \cite{davier} 
  & $-6.4 \pm 1.6$  & $ 8.7\pm 2.4 $\\
\hline
 Ref. \cite{ioffe} 
  & $-6.8\pm 2.1$ & $7\pm 4$ \\
\hline
 Ref. \cite{prades} 
  & $-3.2\pm 2.0 $ & $ -12.4\pm 9.0$ \\
\hline
 Ref. \cite{periscond} 
  & $-9.5\pm 3 $ & $16.2\pm 5$ \\
\hline
Ref. \cite{cirigliano} 
  & $-4.45\pm 0.7$ & $-6.2\pm 3.2$  \\
\hline  
 Ref. \cite{sumrules} 
  & $-4\pm 1 $ & $-1.2\pm 6$ \\
\hline
 This work 
  & $-4\pm 2 $ & $ -12 ~\epm{7}{11} $ \\
\hline
\end{tabular}
\end{center}
\caption{}{Previous extractions of the condensates ordered
  chronologically.
Appropiate rescalings have been performed to allow the comparison of
different extractions.}
\label{lit}
\end{table}

Note that our results agree, at least on the sign, with that
of Refs. \cite{prades},\cite{cirigliano},\cite{sumrules}. 
 This is also true for the higher
dimensional condensates of Ref. \cite{cirigliano2}, 
where the authors 
obtain:
\bea
 \la \mo_{10} \ra= \lp 4.8\pm 1.0 \rp~10^{-2}~\mathrm{GeV}^{10} \ ,
\nonumber
\\
\la \mo_{12} \ra= \lp -1.6 \pm 0.26 \rp~10^{-1}~\mathrm{GeV}^{12} \ .
\eea
in agreement with eq. (\ref{staterr}) , although the errors in our 
determination are somewhat larger.

There are some differences between these previous determinations and the 
present one, eq. (\ref{finalresults}). The first one is that we 
do not make any assumption on the values of the higher dimensional
nonperturbative condensates. In many analysis the effect of
$\la \mo_D \ra$ for $D\ge 10$ is simply neglected to get closed
expressions for the condensates. In our analysis, though, we do not
need to make this hypothesis. Neither we need to assume
that the chiral sum rules are verified, because 
the chiral sum rules enter as constraints in the genetic algorithm training 
(see
fig. (\ref{srstab})). This is  relevant because
previous analysis showed that the chiral sum rules are not verified
 for $s_0=M_{\tau}^2$, except the DMO sum rule, implying that one must
be extremely careful when dealing with them. A second
main difference is the absence of theoretical bias introduced in
other analysis with a choice of a given finite energy
sum rules. Moreover, the smooth interpolating capability of the
neural network lets  the integration range to be taken up to
arbitrarily high energies.

\section{Conclusions }
We have presented a determination of the nonperturbative 
vacuum condensates $\la\mo_6\ra$ and $\la\mo_8\ra$ from the spectral functions
from hadronic tau decays aimed at minimizing the sources of
theoretical bias which might be cause of concern in existing determinations
of these condensates from spectral functions. This
determination is based on  a bias-free neural network parametrization
of the $v_1(s)-a_1(s)$ spectral function, inferred from the data, which
retains all the information on experimental errors and correlations, and
supplemented with the additional theoretical input of the chiral sum rules.

Our final results give 
negative central values for  the dimension 6 and 8 condensates. 
These results take into account the propagation of statistical
errors and their correlations. Morevover, the main source
of systematic error in our procedure is identified as the
choice of relative weights assigned to chiral sum rules in
the fitness function used to train the neural networks. In
the case of the dimension 6 condensate a stability 
analysis can be performed. Higher dimension condensates 
carry larger errors, although the sign of the condensates
seem to remain unaltered.

The sign of the dimension eight condensate $\la \mo_8 \ra$
deserves further comments. Our central value is negative
within statistical errors but is sensitive to the position 
of artificial endpoints added to enforce the vanishing
of the spectral function for large values of $s$. This
produces a systematic bias as possible wiggles of the
spectral function around the asymptotic zero value
are suppreseed. Those wiggles may indeed produce
a change of sign of $\la \mo_8\ra$. This is not the
case in our approach as the smoothness of the neural
network tends to avoid such wiggles, which might lead
to a systematic error. Although our results 
seem to point in the same direction of other recent previous extractions
of $\la \mo_8\ra$ we consider the issue of the sign of the 
vacuum condensate $\la \mo_8\ra$ to remain open.

Another result of this work is the implementation of a  
technique based on genetic algorithm neural network training, which extends
the capabilities of neural network data analysis allowing to
incorporate non-local constraints like convolutions in the training. 
This technique extends previous efforts \cite{structure}
oriented
to the improvement of high energy physics data
analysis, specially for the strong interaction sector.

\vskip 1cm

{\large {\bf Acknowledgments}} 

\vskip 3mm

It is a pleasure to thank S. Peris and E. de Rafael 
for suggesting the application of Ref. \cite{alphas} to this problem
and
for fruitful discussions, and J. Prades and A. Pich for useful
comments and suggestions. We would also 
like to thank A. Hoecker and W. Mader for their help with the experimental
data of the ALEPH and OPAL collaborations respectively.
This work was supported by the projects MCYT FPA2001-3598, 
GC2001SGR-00065  and by the Spanish
 grant AP2002-2415.

\vfill
\eject   

\appendix

\section{Neural network techniques}
\label{nntech}
In this section  a brief review of the neural network
training algorithms that have been used in the this analysis is presented.
Two different learning algorithms have been used 
for the neural network training:
in a first epoch the only contribution to the error function comes from 
the statistical experimental errors, and  the used learning 
algorithm is known as backpropagation. 
Then, in a second training epoch,   
the contribution
from the chiral sum rules is added to the error function. 
As long as convolutions over the neural network
 output are non local constraints, the previous technique is no longer 
useful and another learning algorithm is needed, which is called
genetic algorithms training. Now each 
of this techniques will be introduced, directing 
the interested reader to 
the standard reviews and textbooks \cite{netrev} on
neural networks and applications. 

Learning by backpropagation allows to train multilayer neural 
networks and has proved
to be an excellent tool in classification, interpolation and prediction tasks.
It is a standard technique that has been recently applied to 
data analysis in high energy physics,
 see Ref. \cite{structure}.
The starting point is a set of input-output patters,
\be
\label{pat}
{\left({\bf x}^{\mu},{\bf z}\right)\in R^n\times R^m, \mu=1,\ldots,p}
\ ,
\ee
that  network must learn. In our case, each input-output pattern
consists of a single data point, the input being the energy $s$ and 
the output the spectral function $\rho_{V-A}(s)$.

The basis of the network learning is the error function, also known
as fitness functional.
The error function is defined as the difference between the actual and
the
 desired output
of the net, measured over the training set, and weighted with
the experimental errors. It is given by
\be
\label{errback}
E=\sum_{i=1}^{N_{dat}}\frac{\lp\rho_i^{(exp)}
-\rho_i^{(net)}\rp^2}{\sigma_i^{(exp)^2}} \ .
\ee
Applying the gradient descent minimization procedure, that is, looking for
the direction of steepest descent of the error function,
the appropriate changes in the network parameters such that the
error function decreases can be determined. The error is introduced in the 
units of the last layer by
\be
\label{deltalast}
\Delta_{i}^{(L),\mu}=g'(h_i^{(L)\mu})\lc
o_i(\vec{x}^{\mu})-z_i^{\mu}\rc \ ,
\ee
and then backpropagated to the rest of the network
\be
\Delta_{j}^{(l-1),\mu}=g'(h_j^{(l-1)\mu})\sum_{i=1}^{n_l}\Delta_i^{(l),\mu}
\omega_{ij}^{(l)} \ ,
\ee
and the last step consistes on the update of the weights and thresholds 
of the net
\be
\delta\omega_{ij}^{(l)}=-\eta\sum_{\mu=1}^p\Delta_i^{(l)\mu}\xi_j^{(l-1)\mu}
+\alpha\delta\omega_{ij}^{(l)}(\mathrm{last}) \ ,
\ee
\be
\delta\theta_{i}^{(l)}=-\eta\sum_{\mu=1}^p\Delta_i^{(l)\mu}
+\alpha\delta\theta_{i}^{(l)}(\mathrm{last}) \ ,
\ee
where $\eta$ is the learning rate parameter which controls the velocity
of the training and the term with $\alpha$ is called a momentum term, 
which improves the algorithm so that the training does not get stuck
in a local minimum. The main advantage of this method is that it is
deterministic and it has been used repeatedly in different situations,
always successfully.

As stated above, the second of the neural networks techniques that are used
in our analysis  is called genetic algorithm learning, also known
as natural selection learning.
As long as the chiral sum rules are convolutions of the network output, that is
are non local functions on the error function, 
meaning that they depend not on a single network output but on the
global output, 
the usual backpropagation training techniques are not useful in this
second epoch of our training. Now the error functional
has  the form of eq. (\ref{errback}) but with additional convolutions
of the neural network output
\be
\label{errge}
E=\sum_{i=1}^{N_{dat}}\frac{\lp\rho_i^{(exp)}
-\rho_i^{(net)}\rp^2}{\sigma_i^{(exp)^2}}+\sum_{i=1}^{N_{conv}}w_i\lp
\int_0^{s_0} ds 
f_i\lp \rho^{(net)}(s)\rp-A_i\rp \ ,
\ee
where $A_i$ is the theoretical value of the $i-$th sum rule and $w_i$
is the relative weight of this sum rule. In this case 
genetic algorithms are used, a training method inspired in the evolutionary
theories in biology. In this method, the network parameters
are transformed into bits in an ADN chain. These chains are
replicated, and then some bits are mutated
with a certain probability, and only those chains with the smallest
contribution to the error functional  survive.
By analogy, this method is also known as natural selection learning.
A simple scheme of the recursive process can be seen as follows: 
the starting piont is the set of parameters that define the neural network,
$$ \omega_{11}^{(1)},\omega_{11}^{(2)},\ldots,\theta_1^{(1)},\theta_1^{(2)},
\ldots $$
$$ \Downarrow $$
\bc
Creation of ADN chain
\ec
$$ \Downarrow $$
$$ ADN_1=\left(\omega_{11}^{(1)},
\omega_{11}^{(2)},\ldots,\theta_1^{(1)},\theta_1^{(2)},
\ldots \right)$$
$$ \Downarrow $$
\bc
Replication of ADN chains
\ec
$$ \Downarrow $$
$$ ADN_1=\left(\omega_{11}^{(1)},
\omega_{11}^{(2)},\ldots,\theta_1^{(1)},\theta_1^{(2)},
\ldots \right),
 ADN_2=\left(\omega_{11}^{(1)},
\omega_{11}^{(2)},\ldots,\theta_1^{(1)},\theta_1^{(2)},
\ldots \right)\ldots$$
$$ \Downarrow $$
\bc
Random mutation of bits in the ADN chains
\ec
$$ \Downarrow $$
$$ ADN_1=\left(\omega_{11}^{(1)},
\omega_{11}^{(2)},\ldots,\theta_1^{(1)},\theta_1^{(2)},
\ldots \right),
 ADN_2=\left(\omega_{11}^{(1)}+
\delta_2\omega_{11}^{(1)},\ldots,\theta_1^{(1)}+
\delta_2\theta_1^{(1)},
\ldots \right)\ldots$$
$$ \Downarrow $$
\bc
Selection of best ADN chain
\ec
$$ \Downarrow $$
$$
 ADN_j=\left(\omega_{11}^{(1)}+\delta_j\omega_{11}^{(1)},\ldots,\theta_1^{(1)}+
\delta_j\theta_1^{(1)},
\ldots \right)$$
$$ \Downarrow $$
\bc
New weights and thresholds and decrease of fitness
\ec
$$ \Downarrow $$
$$
\omega_{11}^{(1)}+\delta_j\omega_{11}^{(1)},\ldots,\theta_1^{(1)}+
\delta_j\theta_1^{(1)},
\ldots $$
This is a very simple genetic algorithm, more modifications
could be added to improve its efficiency like crossing between
individuals (characterized by an ADN chain), but our analysis showed
that this
was
not necessary. Its main drawbacks 
are that it is random rather than deterministic as is the 
backpropagation algorithm, and that requires to 
carefully adjust many parameters 
(rate of mutations, size of the population). These parameters have
been adjusted following two requirements: efficiency
of the learning and stability of the result. Learning by
genetic algorithms allows therefore
to impose the theoretical 
constraints from the chiral sum rules in a natural way. 

Finally, we would like to comment on a new learning algorithm that
implements the main advantages of the two methods, that is, it
is deterministic and therefore the simulation time is smaller, but
at the same time it supports non local contributions to the
error function.
During the realization of this work, a novel technique was developed 
 that allowed to use the backpropagation learning algorithms in the
case of eq. (\ref{errge}), when the error functional contains convolutions.
This allowed  to check that the results obtained with the genetic 
algorithm approach were correct. 
In brief, this technique consists in noticing that an integral can be
determined up a any desired accuracy by a finite sum of local 
contributions, when in this context local means that only depends
on one network output. In fact, this is what any numerical integration 
method does, so it is clear that training algorithms for 
backpropagation learning can be implemented. The result is that for
a discretization of the integral of the form
\be
\int dxf(x)=\sum_{j=1}^{n}c_jf_{j} \ ,
\ee
where the coefficients depend on the method, applying the usual 
backpropagation condition (variation of weights and thresholds in the
direction of steepest descent of the error function) to the convolution
term one finds that corresponding equations are the 
backpropagation equations but with eq. (\ref{deltalast}) replaced by
\be
\label{backconv}
\Delta_1^{(L)k}=2\left(\sum_{j=1}^{n_s}c_jf_{b}-A\right)c_a
\frac{df}{do_k}zg'(h_{1,k}^{(L)}) \ ,
\ee
where for simplicity we have only considered one sum rule and
$z$ is a normalization factor present due to the fact that the 
inputs and outputs of the neural network are normalized, so that
the activation function of the neurons are always within
the sensibility range of the activation function.
In eq. (\ref{backconv}) $o_k$ means the output of the network
when the input $x_k$ is introduced, $f(o(x))$ is the convolution
that we want the network to learn and A is its theoretical value.
In this equation each term should be understood as a new pattern for the 
neural network training. From here the usual backpropagation techniques
apply as usual. This novel technique was implemented in the present analysis
but it did not improve neither the quality nor the speed of the
training, 
so the genetic algorithms technique was mantained
for the training with convolutions. This technique, that is called
backpropagation for convolutions, has many applications, and will be
the subject of future work.

\end{document}